\documentclass[10pt,aps,prb,preprintnumbers,amsmath,amssymb,showpacs,notitlepage]{revtex4-1}
\pdfoutput=1
\usepackage{amsmath,braket,mathdots,mathtools,amssymb,xcolor}
\usepackage{graphicx,filecontents}
\usepackage[colorlinks,bookmarks=false,citecolor=blue,linkcolor=red,urlcolor=blue]{hyperref}
\usepackage{lipsum}

\newcommand{\phdag}{\vphantom{\dagger}}

\newcommand{\vect}[1]{\boldsymbol{#1}}
\newcommand{\sech}{\mathrm{sech \,}}

\def\be{\begin{equation}}
\def\ee{\end{equation}}
\def\bea{\begin{eqnarray}}
\def\eea{\end{eqnarray}}
\def\nn{\nonumber\\}
\def\fr#1{(\ref{#1})}
\def\conn#1{\langle\!\langle #1\rangle\!\rangle}

\begin{document}
\title{Self-consistent time-dependent harmonic approximation for the
  sine-Gordon model out of equilibrium}
\author{Yuri D. van Nieuwkerk}%
\author{Fabian H. L. Essler}%
\affiliation{Rudolf Peierls Centre for Theoretical Physics, Parks Road, Oxford OX1 3PU, United Kingdom}

\begin{abstract}
We derive a self-consistent time-dependent harmonic approximation for the quantum
sine-Gordon model out of equilibrium and apply the method to the dynamics
of tunnel-coupled one-dimensional Bose gases. We determine the time
evolution of experimentally relevant observables and in particular
derive results for the probability distribution of subsystem phase
fluctuations. We investigate the regime of validity of the
approximation by applying it to the simpler case of a nonlinear
harmonic oscillator, for which numerically exact results are
available. We complement our self-consistent harmonic approximation
by exact results at the free fermion point of the sine-Gordon model.
\end{abstract}

\date{\today}%
\maketitle
\section{Introduction}
\label{sec:introduction}
The study of isolated quantum many-body systems out of equilibrium has
seen a series of striking successes in the past decades, characterized
by a fruitful interplay between theory and experiment. The possibility
of analyzing the non-equilibrium dynamics of one-dimensional gases in
particular \cite{Ketterle2001,Greiner2001,Kinoshita2004} stimulated a 
multitude of theoretical developments concerning the equilibration of
observables and spreading of correlations and entanglement 
after quantum quenches
\cite{Rigol2007,CalabreseCardy2007,Polkovnikov2011,EFreview,Vidmar2016,DAlessio2016,Gogolin2016,CalabreseCardy2016,CalabreseRev2018}. In
turn, cold atom experiments have been successful in confirming many of
these theoretical ideas
directly\cite{Trotzky2012,Cheneau2012,Gring2012,Langen2013,Langen2015,Kaufman2016}. A
particularly nice example is offered by matter-wave interferometry
\cite{Ketterle1997} using pairs of split one-dimensional Bose gases 
\cite{Schumm2005,Albiez2005,Gati2006,Levy2007,Hofferberth2007,Gring2012,Kuhnert2013,Langen2013,Langen2015},
which can often be modelled theoretically using Luttinger Liquid
theory \cite{Haldane1981,Bistritzer2007}. This approach permits a
theoretical description of the dynamics of observables as well as their
full quantum mechanical probability distribution functions
\cite{Imambekov2007,Kitagawa2010,Kitagawa2011}, which were found to be
in good correspondence with experiment \cite{Gring2012,Langen2015}.  

Of particular interest to our work is the case when a pair of
one-dimensional Bose gases is connected via a finite potential barrier
\cite{Albiez2005,Gati2006,Levy2007,Hofferberth2007}, so that
tunnelling can occur. The low-energy physics of this setup is governed
by a quantum sine-Gordon model \cite{Gritsev2007}
\bea
H_{\mathrm{SG}}&=& H_0 -J \int_{L} dx \cos \phi(x), \label{eq:Sine-Gordon_ham_full}\nn
H_0&=& \frac{v}{2 \pi} \int_{L} dx\; \left[ K (\partial_{x}
\phi(x))^{2} + \frac{1}{K} (\partial_{x} \theta(x))^{2}
\right]. \label{eq:Luttinger_Luiquid_ham} 
\eea
Here the bosonic fields $\phi$ and $\partial_{x}\theta$ satisfy canonical
commutation relations $\left[ \partial_{x}\theta(x) , \phi(y) \right]
= i \pi \delta(x-y)$ and are compactified according to
$\phi=\phi+2 \pi$ and $\theta=\theta+\pi$. The real parameters $v,
J$ and $K>1/2$ are as yet undetermined, but will acquire
physical meaning in what follows. Experiments have focussed on finite
temperature equilibrium properties \cite{Schweigler2017} and
non-equilibrium dynamics in presence of a nonzero initial phase
difference \cite{Pigneur2017} in the large-$K$ regime. The latter
experiments observed damped phase oscillations and relaxation to a
phase-locked state, for which no theoretical explanation is known \cite{Pigneur2018}. 
On the theoretical side there have been a number of works
investigating the dynamics after quantum quenches to the sine-Gordon model. 
The limit  $K\to\infty$ is amenable to a simple harmonic approximation
\cite{Iucci2009,Iucci2010,Foini2015,Foini2017}, while at $K=1/4$ the sine-Gordon
model is equivalent to a free massive Dirac fermion and this can be
used to obtain exact results\cite{Iucci2009,Iucci2010}. In Ref.~\onlinecite{DallaTorre2013} 
a combination of semiclassical and perturbative methods was used to
study the rephasing dynamics for two coherently split condensates
without initial phase difference. Bertini \textit{et al.}~\onlinecite{Bertini2014} investigated
the time dependence of one-point functions in the repulsive regime
$K<1/4$ for quenches from an ``integrable'' initial state by a
combination of quench action\cite{CE2013,CauxQA2016} and linked-cluster
expansion\cite{CEF2012,Schuricht2012} methods. 
In Ref.~\onlinecite{Kormos2016} semiclassical methods\cite{Sachdev1997,Igloi2011} were
applied to the same problem, while quenches from the same class of
initial states to the attractive regime of the sine-Gordon model were
considered in Refs~\onlinecite{Schuricht2017,Horvath2017,Horvath2018a}. 
A novel semiclassical approach was developed in
Ref.~\onlinecite{Pascu2017} and used to determine the time-dependence
of one and two-point functions as well as the probability distribution
of the phase. The truncated conformal space approach\cite{James2018} was applied in
Ref.~\onlinecite{Kukuljan2018} to study the time evolution of two and
four-point functions after a quantum quench. A very recent
work\cite{Horvath2018} addressed the phase-locking behaviour observed 
in the experiments\cite{Pigneur2017} by applying a combination of
numerical methods to the phase dynamics in the sine-Gordon
model. These findings are at variance with the experimental
observations, although the parameter window of the methods does not currently
extend to the relevant regime of weak interactions. This means that in
spite of tentative evidence to the contrary, it is as yet unclear whether
the observed relaxation to a phase-locked state is captured by a
description in terms of a sine-Gordon model.

The aim of this work is to contribute to this discussion by improving
on the known quadratic approximation, valid at weak interaction
strengths, and replacing it by a self-consistent harmonic
approximation which approximates the full cosine potential in a
time-dependent manner. Such an approximation has been successfully
employed for $\phi^{4}$-theory, both in equilibrium \cite{Chang1975}
and out of equilibrium \cite{Sotiriadis2010}, and it has been
formulated for the sine-Gordon model in Ref. \onlinecite{Boyanovsky1998}. We
present an alternative derivation of the method, leading to a set of
coupled nonlinear equations of motion, which we solve numerically. This 
not only yields correlation functions, but also allows for the calculation 
of full distribution functions for the relevant observables. As an application
of this method, we show that for squeezed initial states relevant for cold-atom
experiments, the model exhibits density-phase oscillations with a
time-dependent modulation of the amplitude. This amplitude modulation
depends on the number-squeezing factor which characterizes the initial
state. These results are complemented by exact calculations at the
free fermion point of the sine-Gordon model, where strong damping of
density-phase oscillations is observed.   

\section{Derivation of the self-consistent harmonic approximation} 
\label{sec:sine_gordon_in_self_consistent_approximation}
Our point of departure is the quantum sine-Gordon model
\fr{eq:Luttinger_Luiquid_ham} on a ring of circumference $L$.
We are interested in non-equilibrium dynamics after a quantum quench:
the system is prepared in an initial pure state $|\psi(0)\rangle$ which is
not an eigenstate of $H_{\mathrm{SG}}$ and which satisfies Wick's
theorem. The subsequent time evolution of the system is then described
by the time-dependent Schr\"odinger equation
\be
|\psi(t)\rangle=e^{-iH_{\rm SG} t}|\psi(0)\rangle\ .
\ee
The self-consistent time-dependent harmonic approximation (SCTDHA) consists of
replacing the exact time evolution operator with
\be
e^{-iH_{\rm SG} t}\longrightarrow U_{\rm SCH}(t)=Te^{-i \int_{0}^{t}
  H_{\mathrm{SCH}}(\tau) d \tau}\ , \label{eq:repl_time_evol}
\ee
where
\begin{align}
H_{\mathrm{SCH}}(t)= H_0&-J \int_{L} dx \big[f(x,t) + g(x,t)
  \phi(x) + h(x,t) \phi^{2}(x)\big]. 
\label{eq:replacement_cos}
\end{align}
The time-dependent functions in \fr{eq:replacement_cos} are determined
in a self-consistent way as follows. We assume that the Bose field can
be decomposed into creation/annihilation parts with respect to the
time evolved state  $|\psi_{\rm SCH}(t)\rangle=U_{\rm SCH}(t)|\psi(0)\rangle$
\bea
\phi(x) &=& \braket{\phi(x)}_{t} + \phi^{+}(x,t) + \phi^{-}(x,t)\ ,\nn
\phi^{-}(x,t)|\psi_{\rm SCH}(t)\rangle&=&0=\langle\psi_{\rm
  SCH}(t)|\phi^{+}(x,t)\ ,
\label{decompose}
\eea
where the commutator $[\phi^+(x,t),\phi^-(y,t)]$ is a c-number and
\be
\braket{\phi(x)}_t=\langle\psi_{\rm SCH}(t)|\phi(x)|\psi_{\rm SCH}(t)\rangle.
\ee
The existence of the decomposition \fr{decompose} holds for the class of
initial states described in Appendix \ref{app:Wick}. We then define a
\textit{normal ordering} operation $:\phi^{n}:$ by stipulating that in
a normal ordered expression all $\phi^{-}(t)$ appear on the rightmost side of
any product. In particular we have
\begin{align}
:\phi^{n}: = \sum_{m=0}^{n} \begin{pmatrix}
	n \\
	m
\end{pmatrix} \braket{\phi}_{t}^{n-m} :\left( \phi^{+}(t) + \phi^{-}(t) \right)^{m}:. \label{eq:NO_series}
\end{align}
Applying this normal ordering procedure to $\cos(\phi)$ we find
\be
\cos\big(\phi(x)\big) = \,:\cos\big(\phi(x)\big): 
e^{- \frac{1}{2} \conn{\phi^{2}(x)}_{t}}
=\sum_{n=0}^\infty\frac{(-1)^n}{(2n)!}:\phi^{2n}(x):
e^{- \frac{1}{2} \conn{\phi^{2}(x)}_{t}},
\ee
where $\conn{.}$ denotes connected correlation functions
\be
\conn{\phi^2(x)}_t=\braket{\phi^2(x)}_t-\braket{\phi(x)}_t^2.
\ee
We now use \fr{eq:NO_series} and neglect all higher than quadratic terms in fluctuations
i.e. we set
\begin{align}
:\left( \phi^{+}(t) + \phi^{-}(t) \right)^{m}: \, \longrightarrow 0
  \;\;\; \forall \;\;\; m>2. \label{eq:self_cons_mf} 
\end{align}
This results in the time-dependent Hamiltonian \fr{eq:replacement_cos}
subject to the self-consistency conditions 
\begin{align}
f(x,t) &= \bigg[ \big( 1+ \frac{1}{2} \left( \conn{\phi^{2}(x)}_{t} - \braket{\phi(x)}^{2}_{t} \right) 
  \big)\cos\big(\braket{\phi(x)}_{t}\big)  + 
\braket{\phi(x)}_{t} \sin\big(\braket{\phi(x)}_{t}\big)\bigg]e^{- \frac{1}{2} \conn{\phi^{2}(x)}_{t} }\ , \notag\\
g(x,t) &= \left[ \braket{\phi(x)}_{t}
  \cos\big(\braket{\phi(x)}_{t}\big) -
  \sin\big(\braket{\phi(x)}_{t}\big) \right] e^{- \frac{1}{2} \conn{\phi^{2}(x)}_{t} }\ , \notag \\
h(x,t) &= -\frac{1}{2} \cos \braket{\phi(x)}_{t} e^{- \frac{1}{2}
  \conn{\phi^{2}(x)}_{t} }. 
\label{eq:fg_def}
\end{align}
\subsection{Alternative Derivation}
The SCTDHA is perhaps more naturally derived on the level of the equations of
motion, as is done in Ref.~\onlinecite{Boyanovsky1998}. Since the
cosine term in the sine-Gordon 
Hamiltonian (\ref{eq:Sine-Gordon_ham_full}) contains all positive,
even powers of the field, it generates an infinite set of coupled
partial differential equations relating the time evolution of all
connected $n$-point functions, i.e. a BBGKY-hierarchy. This hierarchy
is truncated by assuming that all connected $n$-point functions are
negligible above a certain order $n$. In the SCTDHA, one truncates at
quadratic order, meaning all higher cumulants are set to zero. For a
Gaussian initial state there will always be some time scale up to
which this is a good approximation, see e.g. Refs
\onlinecite{Bertini2015,Bertini2016}.  Following
Ref.~\onlinecite{Boyanovsky1998} we separate the field into its
expectation value and fluctuations around it 
\be
\phi(x,t)=\langle \phi(x,t)\rangle+\hat{\chi}(x,t).
\ee
The equation of motion of the Bose field is then
\bea
\left( v^{2} \partial_{x}^{2} - \partial_{t}^{2} \right) \phi(x,t) &=&
\frac{v\pi J}{K} \sin\big(\phi(x,t)\big)\nn
&=& \frac{v\pi J}{K} \left[\sin\big(\left\langle \phi(x,t) \right\rangle \big)
\cos\big(\hat{\chi}(x,t)\big)  
+ \cos\big(\left< \phi(x,t) \right> \big) \sin\big(\hat{\chi}(x,t)\big)
\right]\ .
\eea
Assuming that all higher cumulants of the fluctuation field are
negligible the right-hand side can be approximated by
\bea
\frac{v\pi J}{K}\left[ \sin\big(\left\langle \phi(x,t) \right\rangle \big)
:\cos\big(\hat{\chi}(x,t)\big):  
+ \cos\big(\left< \phi(x,t) \right> \big) :\sin\big(\hat{\chi}(x,t)\big):
\right]
 e^{- \frac{1}{2} \conn{\hat{\chi}^{2}}}.
\eea
The equation of motion for the expectation value then becomes
\begin{align}
\left( v^{2} \partial_{x}^{2} - \partial_{t}^{2} \right) \left<
\phi(x,t) \right>  
= \frac{v\pi J}{K} \sin\big(\left< \phi(x,t) \right>\big) e^{- \frac{1}{2}
  \conn{\hat{\chi}^{2}}}.
\label{eq:EOM_classical}
\end{align}
Finally we linearize the equation of motion for the fluctuation field
\begin{align}
\left[ v^{2} \partial_{x}^{2} - \partial_{t}^{2} - \frac{v\pi J}{K}
\cos\big(\left< \phi(x,t) \right>\big)
 e^{- \frac{1}{2} \conn{\hat{\chi}^{2}(x,t)}} \right] \hat{\chi}(x,t)
= 0\ . 
\label{eq:EOM_quadr_quantum}
\end{align}
It is easy to verify that the equations of motion \fr{eq:EOM_classical} and
\fr{eq:EOM_quadr_quantum} are exactly the same as the Heisenberg equations
of motion with regards to $H_{\rm SCH}(t)$
\be
\frac{\partial\phi(x,t)}{\partial t}=iU_{\rm SCH}(t)[H_{\rm
    SCH}(t),\phi(x)]U^\dagger_{\rm SCH}(t).
\ee

\subsection{Mode expansion} 
\label{sub:mode_expansions}
The mode expansions for the Bose field and the dual field can be cast
in the form
\begin{align}
\phi(x) &= \sum_{j} u^{\phdag}_{j} e^{i q_{j} x} \left( b^{\phdag}_{j} - b_{-j}^{\dagger} \right)\ , \label{eq:mode_expansion_phi} \\ 
\frac{\partial_{x} \theta(x)}{\pi} &= \frac{-i}{2 u_{0} L} \left(
b^{\phdag}_{0} + b_{0}^{\dagger} \right) + \sum_{j\neq 0} \frac{i e^{i q_{j} x}}{2u_{j}L} \left( b^{\phdag}_{j} + b_{-j}^{\dagger} \right)\ ,\label{eq:mode_expansion_pi}
\end{align}
where $q_{j} = 2 \pi j/L$ and we have introduced coefficients
\begin{align}
u_{j} &= \begin{cases}
   \Big| \frac{\pi}{2q_{j}LK}\Big|^{1/2}\mathrm{sgn}\left(q_{j}\right), &\text{ if }j \neq 0\ ,\\
  \frac{i}{4} \sqrt{\frac{2v}{K}}&\text{ if }j = 0\ .
\end{cases} \label{eq:u_vect}
\end{align}
The zero momentum modes take account of the periodicity of the dual Bose
 field $\theta(x+L) = \theta(x) +
\pi \delta N$, where $\delta N$ is an
operator with integer eigenvalues \cite{Haldane1981}. In
(\ref{eq:mode_expansion_phi},\ref{eq:mode_expansion_pi}) we have introduced creation/annihilation operators by
\begin{align}
b_{0} = - \left( i\sqrt{\frac{2K}{v}} \phi_{0}+
\frac{1}{2}\sqrt{\frac{v}{2K}} \delta N
\right)\ , \label{eq:ladder_op_zm} 
\end{align}
where $\left[ \delta N, \phi_{0} \right]=i $. The Bose field $\phi(x)$, though compactified with period $2 \pi$ in the original sine-Gordon model, can not be compactified in the SCTDHA, since the harmonic approximation breaks the periodicity of the cosine potential. Instead, we take $\phi(x)$ and its zero mode $\phi_{0}$ to have a spectrum ranging over all of $\mathbb{R}$. Local observables are not affected by this decompactification as long as we consider states where $\braket{\phi(x)}$ lies close to zero, and focus on regimes where $K \gg 1$, such that the variance of $\phi(x)$ is small.

By construction the free part of the Hamiltonian is diagonalized by the above mode expansions, as
\begin{align}
H_{0} = \frac{\pi v (\delta N)^{2}}{2KL} + \sum_{j \neq 0} v |q^{\phdag}_{j}| b_{j}^{\dagger} b_{j}^{\phdag}.
\end{align}
In order to describe time evolution in the self-consistent harmonic
approximation it is convenient to carry out an initial state dependent
canonical transformation.
\subsection{Gaussian initial states} 
\label{sub:gaussian_initial_states}
To guarantee the existence of a time scale over which the SCTDHA
is accurate we prepare our system in a Gaussian
initial state. In the following we restrict ourselves to translationally 
invariant Gaussian pure states for simplicity and refer to
Ref.~\onlinecite{NieuwkerkPrep} for a discussion of the general case. 
In terms of the bosonic creation and annihilation operators any
translationally invariant Gaussian pure state can be written in the 
form 
\begin{align}
\ket{V,\vartheta,\varphi} = \exp \left( V^{\phdag}_{0} \,\sech \vartheta^{\phdag}_{0}\, b^{\dagger}_{0}+\frac{1}{2}\sum^{\phdag}_{k} e^{i \varphi_{k}} \tanh \vartheta^{\phdag}_{k} \, b^{\dagger}_{k} b^{\dagger}_{-k} \right) \ket{0}, \label{eq:init_state}
\end{align} 
where $\vartheta_{k} = \vartheta_{-k}$ and $\varphi_{k} =
\varphi_{-k}$ are real coefficients. To simplify some of the equations
below we introduce
\begin{align}
V_{k} = \delta_{k,0} V_{0} \in \mathbb{C}\ .
\end{align}
The operators
\begin{align}
a^{\phdag}_{k} = \cosh \vartheta^{\phdag}_{k} \,b^{\phdag}_{k} -
e^{i\varphi_{k}}\,\sinh^{\phdag} \vartheta^{\phdag}_{k} \,
b^{\dagger}_{-k} - V^{\phdag}_{k}\ , \label{eq:app:a_as_bs}
\end{align}
annihilate the initial state
\be
a_k\ket{V,\vartheta,\varphi}=0\ .
\label{anni}
\ee
The two sets of creation and annihilation operators are related by a
canonical transformation
\begin{align}
b^{\phdag}_{k} =  \cosh \vartheta^{\phdag}_{k} \left[ a^{\phdag}_{k} +
  V^{\phdag}_{k} \right]  + e^{i\varphi_{k}}\,\sinh
\vartheta^{\phdag}_{k} \left[ a^{\dagger}_{-k} + V^{\dagger}_{-k}  \right]. 
\label{eq:b_as_as} 
\end{align}
\subsection{Equations of motion} 
\label{sub:equations_of_motion}
The Hamiltonian $H_{\mathrm{SCH}}(t)$ has a mode expansion of the form
\be
H_{\mathrm{SCH}}(t) = \sum_{j}\left[b^{\dagger}_{j} A^{\phdag}_{j}(t) b^{\phdag}_{j} +
\frac{1}{2} \big(
b^{\phdag}_{j} B^{*\phdag}_{j}(t) b^{\phdag}_{-j} +{\rm h.c.}\big)\right]
 +D(t) \left( b^{\phdag}_{0} - b_{0}^{\dagger} \right) + C(t)\ , 
\label{eq:generic_quadratic_ham}
\ee
where the coefficients $A_{j}(t), B_{j}(t)$ and $D(t)$ are functions of $g(t)$ and $h(t)$ via 
\begin{align}
A_{j}(t) &= v |p_{j}| - 2 J L |u_{j}|^{2} h(t)\ , \notag \\
B_{j}(t) &= v |p_{0}| \delta_{j0} + 2 J L |u_{j}|^{2} h(t)\ , \label{eq:ABD}\\
D(t) &= -J L u_{0} g(t)\ .\notag
\end{align}
In the above, we have defined $p_{0} = 2 \pi / v L$, and $C(t)$ is a
real scalar which does not affect the equations of motion. The
functions $g(t)$ and $h(t)$ are position independent as we have
imposed periodic boundary conditions and assumed the initial state to
be translationally invariant. The time evolution of $b_{j}$-operators
is obtained from the Heisenberg equation of motion 
\begin{align}
    i \frac{d}{dt} b_j(t) = U_{\rm SCH}(t)\left[ b_j,
      H_{\mathrm{SCH}}(t)\right]U^\dagger_{\rm SCH}(t).
\label{eq:Heisenberg_EOM}
\end{align}
As $H_{\rm SCH}(t)$ couples only modes with either the same or equal
but opposite index and in view of \fr{eq:b_as_as} the time evolved annihilation operators
can be expressed as 
\begin{align}
b^{\phdag}_{j}(t) = \delta_{j,0} R^{\phdag}(t) + S^{\phdag}_{j}(t) a^{\phdag}_{j} + T_{j}^{*\phdag}(t) a_{-j}^{\dagger}\ .
\label{eq:Ansatz}
\end{align}
The initial conditions follow from \fr{eq:b_as_as}
\begin{align}
R^{\phdag}(0) &= V^{\phdag}_{0} \cosh \vartheta^{\phdag}_{0} + V^{*\phdag}_{0} e^{i \varphi_{0}} \sinh \vartheta^{\phdag}_{0}\ ,\quad
S^{\phdag}_{j}(0) = \cosh \vartheta^{\phdag}_{j}\ ,\quad
T_{j}^{*\phdag}(0) = e^{i \varphi_{j}} \sinh \vartheta^{\phdag}_{j}\ .
\end{align}
The time dependence of $R(t), S_j(t)$ and $T_j(t)$ is obtained by
substituting \fr{eq:Ansatz} in to \fr{eq:Heisenberg_EOM}, which gives a
system of coupled, first order differential equations
\bea
i\dot{R}(t) &=& A_{0}(t)R(t) + B_{0}(t) R^{*}(t) - D(t)\ , \nn
i\dot{S}_{j}(t) &=& A_{j}(t)S_{j}(t) + B_{j}(t) T_{-j}(t)\ , \label{eq:ODE_sys_generic} \nn
-i\dot{T}_{j}^{\phdag}(t) &=& A^{*}_{j}(t)T^{\phdag}_{j}(t) +
B^{*}_{j}(t) S_{-j}^{\phdag}(t).
\eea
We stress that Eqns (\ref{eq:ODE_sys_generic}) are \textit{non-linear}
as $A$, $B$ and $D$ are functions of $R$, $S$ and $T$ by virtue of the
self-consistency conditions \fr{eq:fg_def}. The time evolved Bose
fields in our SCTDHA are given by
\begin{align}
\phi(x,t) &= -2 |u_{0}| \mathrm{Im} \left(R(t)\right) + \sum_{j} u_{j}
e^{i q_{j} x} \left( Q^{\phdag}_{j}(t) a^{\phdag}_{j} -
Q^{*\phdag}_{-j}(t) a^{\dagger}_{-j}
\right)\ ,\label{eq:phi_time_evol} 
\end{align}
where we have defined
\begin{align}
Q^{\phdag}_{j}(t) = S^{\phdag}_{j}(t) - T^{\phdag}_{-j}(t), \;\;\;\;
\bar{Q}^{\phdag}_{j}(t) = S^{\phdag}_{j}(t) +
T^{\phdag}_{-j}(t)\ . \label{eq:def_Q} 
\end{align}
Using that $a_{j} \ket{V,\vartheta,\varphi} = 0$ it is then
straightforward to obtain equal-time correlation functions of the Bose field
\begin{align}
\left< \phi(x,t) \right> &= -2 |u_{0}| \mathrm{Im} \left(R(t)\right)\ \label{eq:phi_exp_val},\\
\left< \phi(x,t) \phi(y,t) \right>_{\rm conn} &= \sum_{j} |u_{j}|^{2}
|Q_{j}(t)|^{2} \cos\big( q_{j} \left( x-y \right)
\big). \label{eq:two_point_connected_transpose} 
\end{align}
These expectation values determine the functions $g(t)$, $h(t)$ and by \fr{eq:ABD} the parameters $A_{j}(t)$,
$B_{j}(t)$, $D_j(t)$. Substituting back into
(\ref{eq:ODE_sys_generic}) we arrive at a closed system of
differential equations for $R_j(t)$, $S_{j}(t)$ and $T_{j}(t)$.
We solve this nonlinear system numerically to obtain the full
time evolution of local operators in our SCTDHA.

\subsection{Full distribution functions and multipoint correlation functions} 
\label{sub:full_distribution_functions_and_multipoint_correlation_functions}
A nice feature of the SCTDHA is that it makes it possible to analyze
not only expectation values of local operators, but the full quantum
mechanical probability distributions of observables on
subsystems. This is of considerable experimental and theoretical
interest \cite{Imambekov2007,Smith2013,Schweigler2017,cd-07,lp-08,ia-13,sk-13,e-13,k-14,sp-17,MC2016,CoEG17,nr-17,lddz-15,Nieuwkerk2018,BastianelloFCS2018,Groha18,BastianelloShG2018}. An example
relevant to realizations of the sine-Gordon model in split one-dimensional
Bose gases are the probability distributions for the real
and imaginary parts of the operator \cite{Imambekov2007,Kitagawa2010,Kitagawa2011} 
\begin{align}
\hat{\mathcal{O}}_{\ell} = \int_{-\ell/2}^{\ell/2} dx\  e^{i
  \hat{\phi}(x)}. \label{eq:FCS_complex_operator} 
\end{align}
It is convenient to define a joint probability distribution of the
commuting operators ${\rm Re}(\hat{\mathcal{O}}_{\ell})$ and ${\rm
  Im}(\hat{\mathcal{O}}_{\ell})$ 
\bea
F_\ell(t,a,b)=\langle\psi_{\rm SCH}(t)|\delta\big({\rm Re}(\hat{\mathcal{O}}_\ell)-a\big)
\delta\big({\rm Im}(\hat{\mathcal{O}}_\ell)-b\big)|\psi_{\rm SCH}(t)\rangle\ .
\eea
As shown in Appendix
\ref{sec:joint_distribution_functions_for_the_phase_operator} it is
possible to obtain a multiple integral representation for this
quantity in the framework of the SCTDHA
\bea
F_{\ell}(t,a,b) &=& \int_{-\infty}^\infty\prod_j\left[
d\alpha_j d\beta_j\frac{e^{- \frac{1}{2}\left|
Q_{j}(t) \right|^{-2} \left( \alpha^{2}_{j} + \beta^{2}_{j}
  \right)}}{2\pi \left| Q_{j}(t) \right|^{2}}\right]
\delta\bigg(a - \int_{-\ell/2}^{\ell/2} dx \cos \left(
\Phi(x,t,\boldsymbol{\alpha}, \boldsymbol{\beta}) \right)\bigg)\nn
&&\qquad\qquad\times\ \delta \bigg(b - \int_{-\ell/2}^{\ell/2} dx \sin \left( \Phi(x,t,\boldsymbol{\alpha}, \boldsymbol{\beta}) \right) \bigg)\ , \label{eq:FDF}
\eea
where
\begin{align}
\Phi(x,t,\boldsymbol{\alpha},\boldsymbol{\beta}) =  \left< \phi(0,t) \right> - \sum_{j} |u_{j}| \Big( \alpha_{j} \cos \left( p_{j} x \right) + \beta_{j} \sin \left( p_{j} x \right) \Big). \label{eq:phi_fun_general}
\end{align}
We see that the distribution function is determined by the expectation
value $\langle \phi(0,t) \rangle$, set by $R(t)$ via
(\ref{eq:phi_exp_val}), along with quadratic fluctuations $\alpha_{j}$
and $\beta_{j}$, determined by the covariance matrix $\left| Q_{j}(t)
\right|^{2}$. The essential quantities $R(t)$ and $Q(t)$ are obtained
by solving the nonlinear, self-consistent system of equations
(\ref{eq:ODE_sys_generic}). The distribution function (\ref{eq:FDF})
can be conveniently sampled: one draws numbers $\alpha_{j}$ and
$\beta_{j}$ from a Gaussian distribution with covariance matrix
$\left| Q_{j}(t) \right|^{2}$ and computes the corresponding values of
$ \int_{-\ell/2}^{\ell/2} dx \exp \left(i
\Phi(x,t,\boldsymbol{\alpha},\boldsymbol{\beta}) \right)$. Placing 
real and imaginary parts of these values in a two-dimensional
histogram and normalizing the result yields $F_{\ell}(t,a,b)$. 
Examples of such distribution functions are presented in 
Section \ref{sub:time_evolution_with_the_full_self_consistent_harmonic_Hamiltonian}. As a
corollary of the derivation in Appendix
\ref{sec:joint_distribution_functions_for_the_phase_operator}, we also
obtain multi-point correlation functions of the vertex operator $e^{i
  \sigma \phi(x)}$, e.g. 
\begin{align}
\left< e^{i \sigma \phi(x,t)} e^{i \tau \phi(0,t)} \right> = e^{i (\sigma + \tau) \left< \phi(0,t) \right> } e^{- \frac{1}{2} \sum_{j} |u_{j}|^{2}|Q_{j}(t)|^{2} \left( \sigma^{2} + \tau^{2} + 2 \sigma \tau \cos \left( q_{j} x \right)  \right) }. \label{eq:phase_vertex_2pt}
\end{align}


\section{Self-consistent harmonic approximation in equilibrium} 
\label{sub:linear_response_above_the_ground_state_}
If we choose self-consistent normal ordering with respect to the
ground state rather than some time evolved initial state our
approximation reduces to the usual self-consistent harmonic
approximation for the sine-Gordon model \cite{Sakaguchi1982}. In the linear response regime at zero temperature 
many exact results are available for the sine-Gordon model, see e.g.
Ref.~\onlinecite{EsslerKonik2005}, and it is instructive to use these
to benchmark the SCHA. The exact breather mass of the sine-Gordon
model is\cite{Zamo1995}  
\begin{align}
\Delta_{1} = 2 \sin \left( \frac{\pi \chi}{2} \right) \frac{2}{\sqrt{\pi}} \frac{v}{\xi} \frac{\Gamma(\chi/2)}{\Gamma((1+\chi)/2)} \left[ \frac{\pi}{2} \frac{\xi^{2}}{v} J \frac{\Gamma(\frac{1}{1+\chi})}{\Gamma(\frac{\chi}{1+\chi})} \right]^{(1+\chi)/2}, \label{eq:gap_ex}
\end{align}
where $\chi = 1/(8K-1)$ and the length scale $\xi$ corresponds to a
cutoff in momentum space at $k_{c} = 2 \pi /\xi$. 

\begin{figure}[htbp]
\centering
\includegraphics[width=0.5\textwidth]{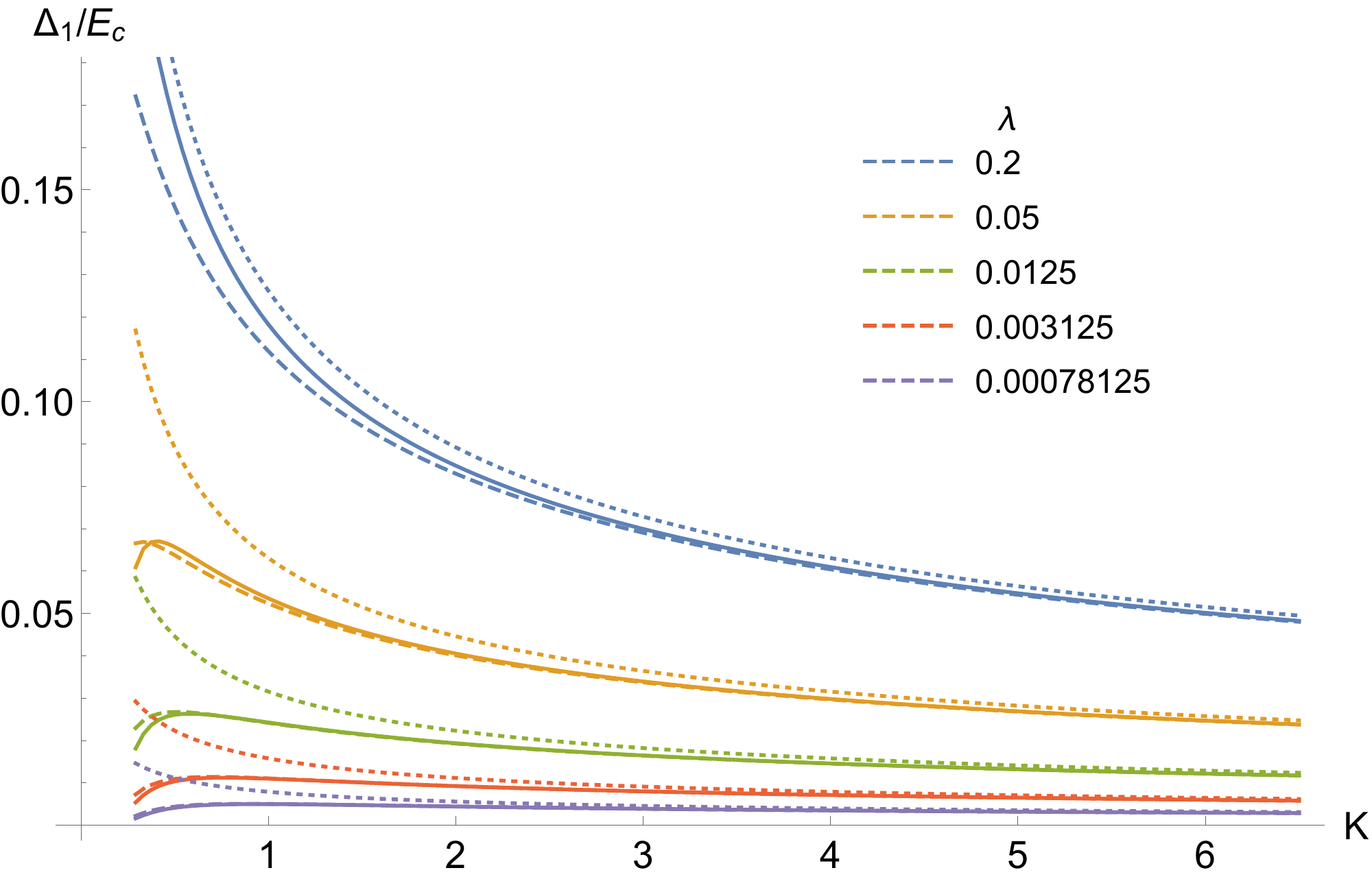}
\caption{Comparison between the mass gap for the fully
         quadratic model with $h = -1/2$ (dotted), the SCHA (dashed)
          and the exact result for the sine-Gordon Hamiltonian (solid
          curves), for several values of the dimensionless coupling $\lambda = \xi^{2} J/v$. The gaps are plotted via their ratio with
          the cutoff energy scale, $ E_{c} = 2 \pi v/\xi$.} 
\label{fig:GS_gaps}
\end{figure}

Normal ordering with regards to the (self-consistent) ground state
results in a time-independent Hamiltonian of the same structure as
$H_{\mathrm{SCH}}(t)$ in (\ref{eq:generic_quadratic_ham}) and
(\ref{eq:ABD}), but with time-independent parameters
\be
g=0\ ,\qquad h=- \frac{1}{2} e^{- \frac{1}{2} \braket{\phi^{2}}}.\label{eq:app_h_GS}
\ee
This Hamiltonian can be diagonalized by a Bogoliubov transformation of the
$b$-operators
\bea
b^{\phdag}_{j}&=&\cosh(\gamma_{j})c^{\phdag}_{j} + \sinh(\gamma_{j})
c^{\dagger}_{-j}\ ,
\label{eq:mass_quench_bogoliubov_shifted}\nn
e^{-2 \gamma_{j}} &=& \frac{\pi}{2KL|u_{j}|^{2}}[\left( v q_{j}
  \right)^{2} + \Delta^{2}]^{-\frac{1}{2}}\ ,
\eea
where we have defined 
\begin{align}
\Delta^{2}= -2h \frac{\pi v J}{K}.
\end{align}
In terms of the Bogoliubov bosons we have
\begin{align}
H_{\mathrm{SCH}}= \sum_{j}\left[ \sqrt{\left( v q_{j} \right)^{2} +
  \Delta^{2}}\, c^{\dagger}_{j} c^{\phdag}_{j}\right] + \tilde{C}\ .
\end{align} 
The ground state of $H_{\rm SCH}$ is the vacuum state of the
$c$-bosons $c_j|0\rangle=0$. The self-consistency condition for $h$ is
then obtained by calculating $\langle\phi^2\rangle=\langle 0|\phi^2(x)|0\rangle$
\be
\langle \phi^2\rangle=
 \frac{\pi v}{2KL} \sum_{j} \frac{1}{\sqrt{v^{2} q_{j}^{2} + \frac{\pi vJ}{K} e^{- \frac{1}{2} \braket{\phi^{2}}}}}.
\ee
A simple quadratic (\textit{non} self-consistent) approximation of
$H_{\mathrm{SG}}(t)$ \cite{Foini2015,Foini2017} would be given by $g
= 0$ and $h = -1/2$, so that 
\begin{align}
\Delta_{\mathrm{qdr}}^{2} = \frac{\pi v J}{K}.
\label{eq:gap_quadr}
\end{align}
In Fig.~\ref{fig:GS_gaps}, we present a comparison between the gap of
the first breather in the sine-Gordon model (solid lines), the gap in
the completely quadratic model (dotted lines) and the gap in the SCHA
(dashed line). This is the appropriate comparison to make because in
the $K$ regime of interest the first breather has the smallest
excitation gap over the ground state. For large enough values of $K$,
both the SCHA and the fully quadratic model provide accurate
approximations. For smaller values of $K$, however, the
self-consistent approximation clearly offers a much better prediction
of $\Delta$ than the simple harmonic approximation does. Close to the
Luther-Emery point, which lies at $K=1/4$ in our conventions, the
predictions from the SCHA become poor as well.  


\section{Realization by tunnel-coupled Bose gases} 
\label{sec:Realization_by_tunnel-coupled_bose_gases}
A very interesting experimental realization of the sine-Gordon
Hamiltonian (\ref{eq:Sine-Gordon_ham_full}) arises by tunnel-coupling
a pair of one-dimensional Bose gases\cite{Gritsev2007,Schweigler2017}
$H=H_{\rm LL}+H_{\rm tunn}$ with
\bea
H_{\rm LL} &=& \sum_{j=1,2}\int dx
\Bigg[  \frac{1}{2m} \partial_{x} \psi_{j}^{\dagger} \partial_{x} \psi_{j}^{\phdag} + g\, \psi_{j}^{\dagger} \psi_{j}^{\dagger} \psi_{j}^{\phdag} \psi_{j}^{\phdag} \Bigg]\ ,
\label{eq:Lieb-Liniger}\\
H_{\rm tunn} &=& - T_{\perp} \int dx \left( \psi_{1}^{\dagger} \psi_{2}^{\phdag} + \psi_{2}^{\dagger} \psi_{1}^{\phdag} \right). \label{eq:micr_tunneling}
\eea
Here $\psi_j$ are complex Bose fields with commutation relations
$\left[\psi_{i}(x),\psi^{\dagger}_{j}(x^{\prime})\right] =
\delta_{i,j}\delta(x-x^{\prime})$. At low energies the model
\fr{eq:Lieb-Liniger}, \fr{eq:micr_tunneling} can be bosonized
using\cite{Haldane1981}
\be
\psi_{j} \sim \sqrt{\rho_{0} + \partial_{x} \theta_{j}(x) /\pi}
\;e^{i \phi_{j}(x)}\ ,
\ee
where $\phi_{j}$ are real Bose fields, $\theta_{j}$ the associated
dual fields and $\rho_0$ the average density of bosons. Expressing the
resulting Hamiltonian in terms of symmetric and antisymmetric
combinations $\phi_{s,a} = \phi_{1} \pm \phi_{2}$, $\theta_{s,a} = \left(
\theta_{1} \pm \theta_{2} \right)/2$ gives a decoupled theory of a
free compact Boson and a sine-Gordon model 
\begin{align}
{\cal H} = \sum_{j=s,a} \frac{v}{2 \pi} \int dx\; \left[K
  (\partial_{x} {\phi}_{j}(x))^{2} + \frac{1}{K} 
  (\partial_{x} {\theta}_{j}(x))^{2} \right] -J
\int dx\ \cos\hat{\phi}_a(x)\ .
\label{eq:Ham_Bosonized}
\end{align}
The (less relevant) coupling between the two sectors will be analyzed in
a forthcoming publication\cite{NieuwkerkPrep}. Importantly the
symmetric sector only gives negligible contributions to the
experimentally relevant observables in the large-$K$
regime\cite{Nieuwkerk2018}. As the initial states of interest do not
mix the two sectors and the observables of interest only involve $\phi_a$
it is possible to restrict the analysis to the sine-Gordon model
describing the antisymmetric sector. To ease notation we drop the
corresponding subscript in what follows.

The cutoff for the low-energy description \fr{eq:Ham_Bosonized} is
given by the healing length of the gas $\xi = \pi / m v$ and we have defined
$J = 2\rho_{0} T_{\perp}$. For weak
interactions, the effective parameters $v$ and $K$ can be related
\cite{Cazalilla2004,Gritsev2007} to the parameters in the microscopic
model (\ref{eq:Lieb-Liniger}) by
\begin{align}
v &= \frac{\rho_{0}}{m} \sqrt{\gamma} \left( 1 -
\frac{\sqrt{\gamma}}{2 \pi}\right)^{1/2}, \quad K =
\frac{\pi}{2\sqrt{\gamma}} \left( 1 - \frac{\sqrt{\gamma}}{2
  \pi}\right)^{-1/2}. \label{eq:micr_pars} \notag 
\end{align}
Here $\gamma = m g/\rho_{0}$ is the dimensionless interaction
parameter. For later convenience we define a dimensionless coupling
constant for the cosine term by 
\be
\lambda = \frac{\xi^{2} J}{v} .
\label{eq:dimensionless_coupling}
\ee

In the experiments by the Vienna group an initial state is prepared
by splitting of a single one-dimensional condensate into
two\cite{Pigneur2017}, which can be modelled by an initial condition
\cite{Kitagawa2010,Kitagawa2011}
\begin{align}
    \left< \frac{\partial_{x} \hat{\theta}(x)}{\pi}\frac{\partial_{y} \hat{\theta}(y)}{\pi} \right>_{c} &= \eta \frac{\rho}{2} \delta_{\xi}(x-y)\ .
\end{align}
Here $\delta_{\xi}$ denotes a delta function which is smeared over
the healing length of the gas. In terms of the squeezed coherent state
(\ref{eq:init_state}), this initial condition is
obtained\cite{Kitagawa2011} by choosing Bogoliubov angles $\varphi_{j}
= 0$ and  
\begin{align}
e^{-2\vartheta_{j}} = \begin{cases}
  \frac{|q_{j}|K}{\pi \eta \rho}\ , &\text{ if } j \neq 0\ ,\\
  \frac{4K}{v L \eta \rho}\ , &\text{ if } j = 0\ .
\end{cases}\label{eq:alpha_def}
\end{align}
The parameter $\eta$ tunes the number and phase fluctuations in the initial state. 

\subsection{Choice of parameters} 
\label{sec:note_on_the_choice_of_parameters}

In order to enable a comparison with experimental observations the
parameters defining our model \fr{eq:Ham_Bosonized} should be fixed
following Ref.~\onlinecite{Kuhnert2013}: the one-dimensional density
is taken to be $\rho_{0} = 45 \, \mu \mathrm{m}^{-1}$, the healing
length $\xi= \hbar\pi / mv = \pi \times 0.42 \,\mu \mathrm{m}$ and 
longitudinal size $L = 160 \, \xi$. Note that the latter is a factor
$2$ larger than the length reported in \onlinecite{Kuhnert2013}. We have
made this adjustment to be able to follow the dynamics over longer
timescales, before boundary effects come into play. For the case of
$^{87}\mathrm{Rb}$ atoms, the above amounts to $L \approx 212 \,
\mathrm{\mu m}$, with a sound velocity given by $v \approx 1.738 \cdot
10^{-3} \, \mathrm{m}/\mathrm{s}$ and a Luttinger parameter of $K
\approx 28$, in our conventions. 

In order to explore the SCTDHA more generally we have also considered
smaller values of the Luttinger parameter $K$. In
Figs~\ref{fig:QM_nu}, \ref{fig:QM_nu_4}, \ref{fig:QM_nu_8} and
\ref{fig:envelope_increase} we show results for $K=1$, where the
difference between the SCTDHA and the simple harmonic approximation is
much larger. The free fermion results shown in
Fig.~\ref{fig:Dens_Phase_Luther_Enery} correspond to $K=1/4$.

\subsection{Time-evolution of the zero mode} 
\label{sub:time_evolution_of_the_zero_mode}
As we have restricted our analysis to translationally invariant
situations the zero momentum modes of the Bose fields play a key role. In the 
full Hamiltonian \fr{eq:generic_quadratic_ham} the zero momentum
modes are sensitive to the dynamics of the finite momentum modes by
virtue of the self-consistency conditions. It is instructive to ignore
such effects and consider the SCTDHA for a toy model that involves
only the zero mode 
\begin{align}
H_{\mathrm{J}} = \frac{\pi v}{2 KL} \delta \hat{N}^{2} - J L
\cos\big(\hat{\phi}_{0}\big), \label{eq:HJ} 
\end{align}
where $\left[ \delta \hat{N}, \hat{\phi}_{0} \right] = i$ and we have
retained the various parameters from the full model. As
\fr{eq:HJ} involves only a single degree of freedom it is
straightforward to obtain exact results by numerically integrating the
corresponding Schr\"{o}dinger equation. This allows us to benchmark
the SCTDHA. As initial state we choose a squeezed state $|\chi(0)\rangle$
with wave function in the $\phi$-representation
\begin{align}
\chi(\phi) = \left( \frac{1}{2\pi \sigma^{2}} \right)^{1/4}
e^{-\frac{\left( \phi - \Phi_{0} \right)^{2}}{4\sigma^{2}} } e^{-i
  \delta N_{0} \phi}, \label{eq:QM_WF_phi} 
\end{align}
where $\sigma^{2} = 1/\left( 2 \eta \rho L \right) $ and $\Phi_0$,
$\delta N_0$ are free parameters. In the SCTDHA the Hamiltonian
\fr{eq:HJ} is replaced by
\be
H_{\mathrm{J}}^{\prime} = \frac{\pi v}{2 KL} \delta \hat{N}^{2} -
J L \left( f(t) + g(t) \hat{\phi}_{0} + h(t)
\hat{\phi}_{0}^{2} \right). \label{eq:HJ_prime}
\ee
The self-consistency conditions for $f$, $g$ and $h$
are obtained from (\ref{eq:fg_def}) by replacing $\phi(x) \rightarrow \hat{\phi}_{0}$. For
reference, we also consider time evolution with a simple harmonic
Hamiltonian obtained from \fr{eq:HJ} by expanding the cosine to second
order in $\hat{\phi}_0$
\begin{align}
H_{\mathrm{HO}} = \frac{\pi v}{2 KL} \delta \hat{N}^{2} + \frac{J
  L}{2} \hat{\phi}_{0}^{2},\label{eq:HHO} 
\end{align}
The ground state wave function of $H_{\mathrm{HO}}$ is given by
(\ref{eq:QM_WF_phi}) with $\Phi_{0}=0=\delta N_{0}$ and
$\eta_{0} \equiv \rho^{-1}\sqrt{JK/v \pi}$.

\begin{figure}[ht]
\centering
(a)\ \includegraphics[width=0.4\textwidth]{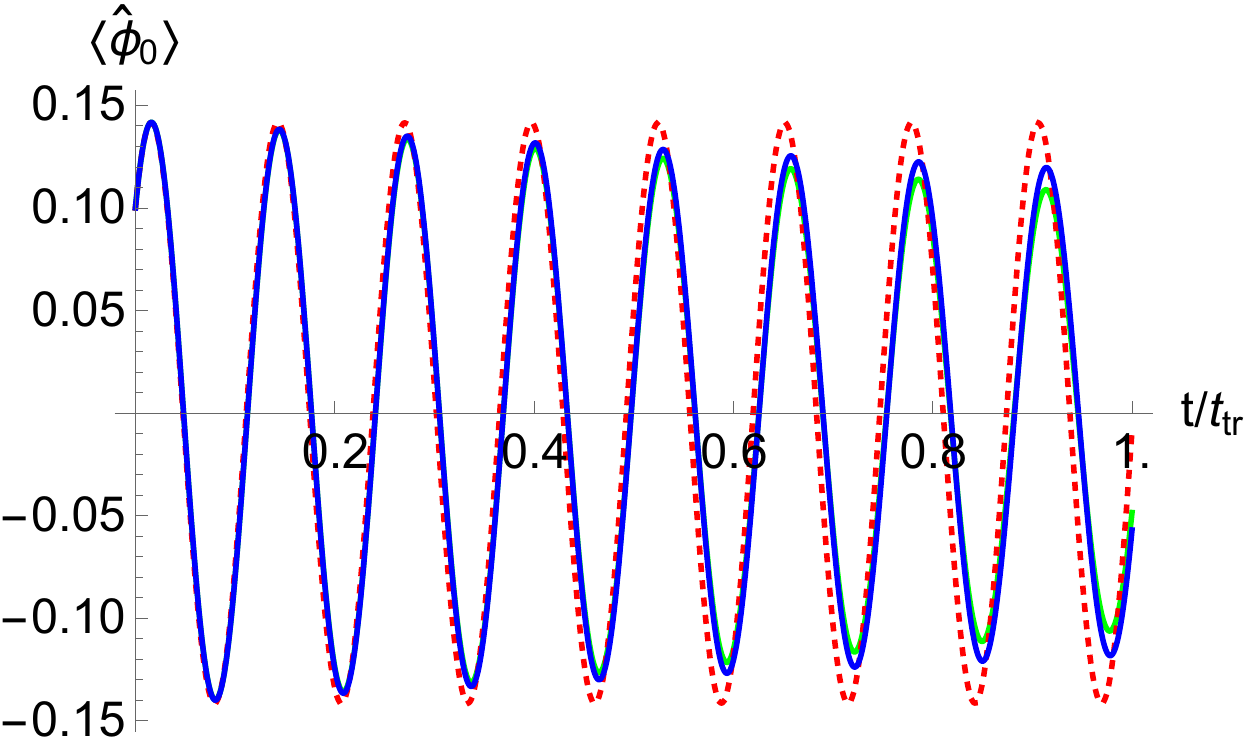}\quad
(b)\ \includegraphics[width=0.4\textwidth]{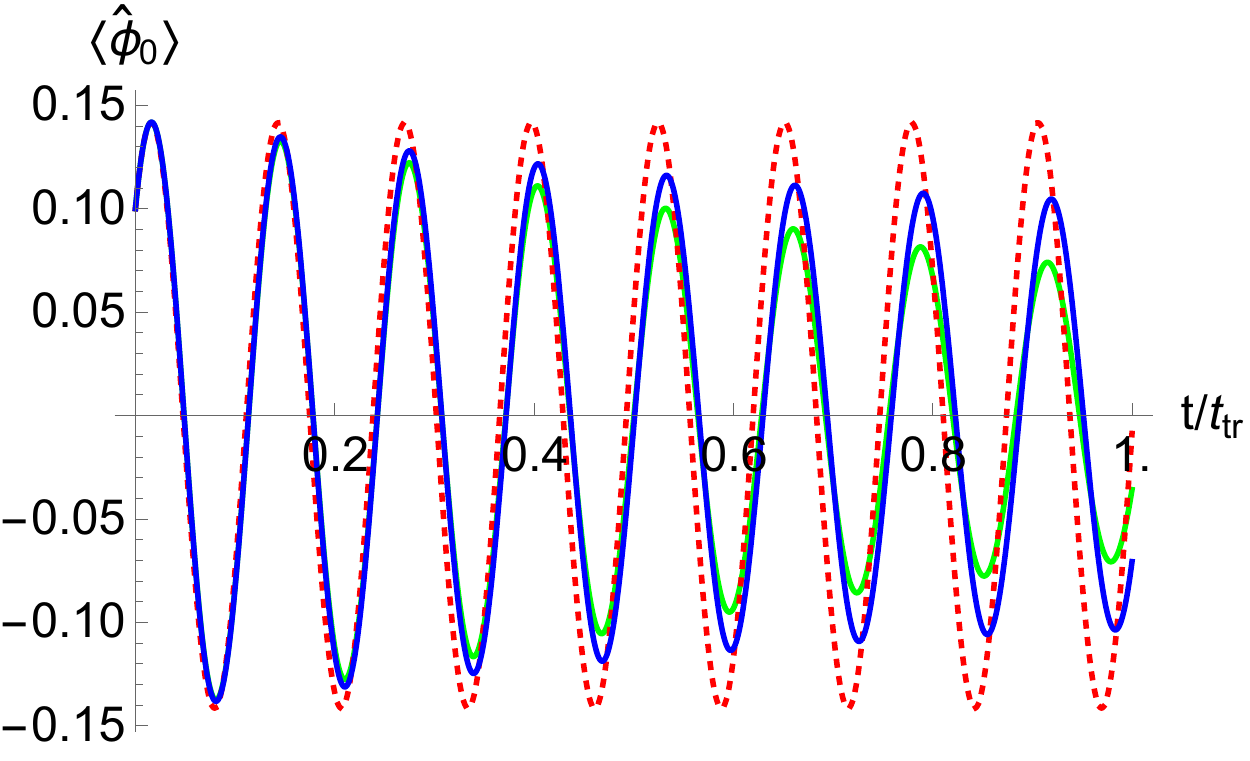}
\caption{Time-evolution of $\left< \hat{\phi}_{0} \right> $ under the full
Hamiltonian $H_{\rm J}$ (green line), the quadratic approximation
$H_{\rm HO}$ (red dots) and the self-consistent harmonic
approximation $H'_{\rm J}$ (blue line). The parameters are as
described in section \ref{sec:note_on_the_choice_of_parameters} and
$\lambda = 0.12$, $K=1$ and (a) $\eta=4 \eta_{0}$; (b) $\eta=8
\eta_{0}$. Times are displayed in units of the ``traversal time''
$t_{\mathrm{tr}} = L/(2v)$ \cite{EFreview}. We have chosen the value
$K=1$ to highlight the differences between the three results, which
are more pronounced for small $K$. Increasing the value of K leads to
a better agreement between the three lines.}
\label{fig:QM_nu}
\end{figure}

\begin{figure}[ht]
\centering
(a)\ \includegraphics[width=0.4\textwidth]{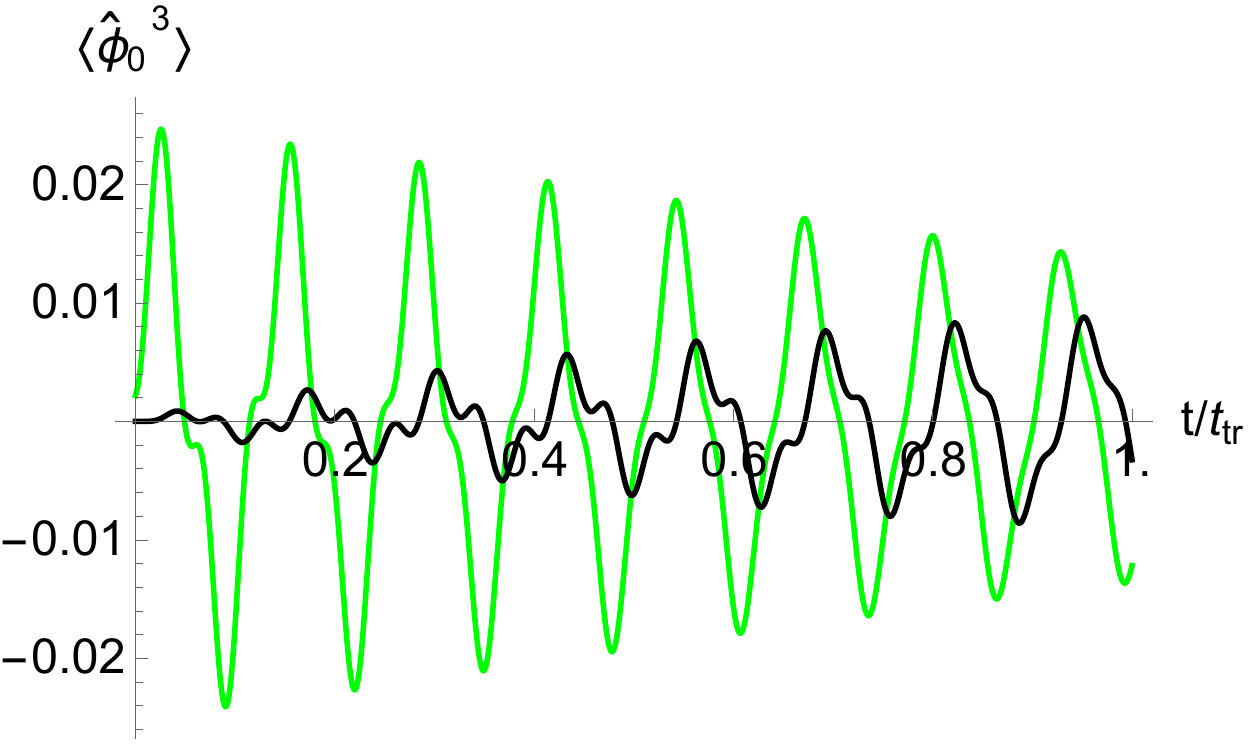}\quad
(b)\ \includegraphics[width=0.4\textwidth]{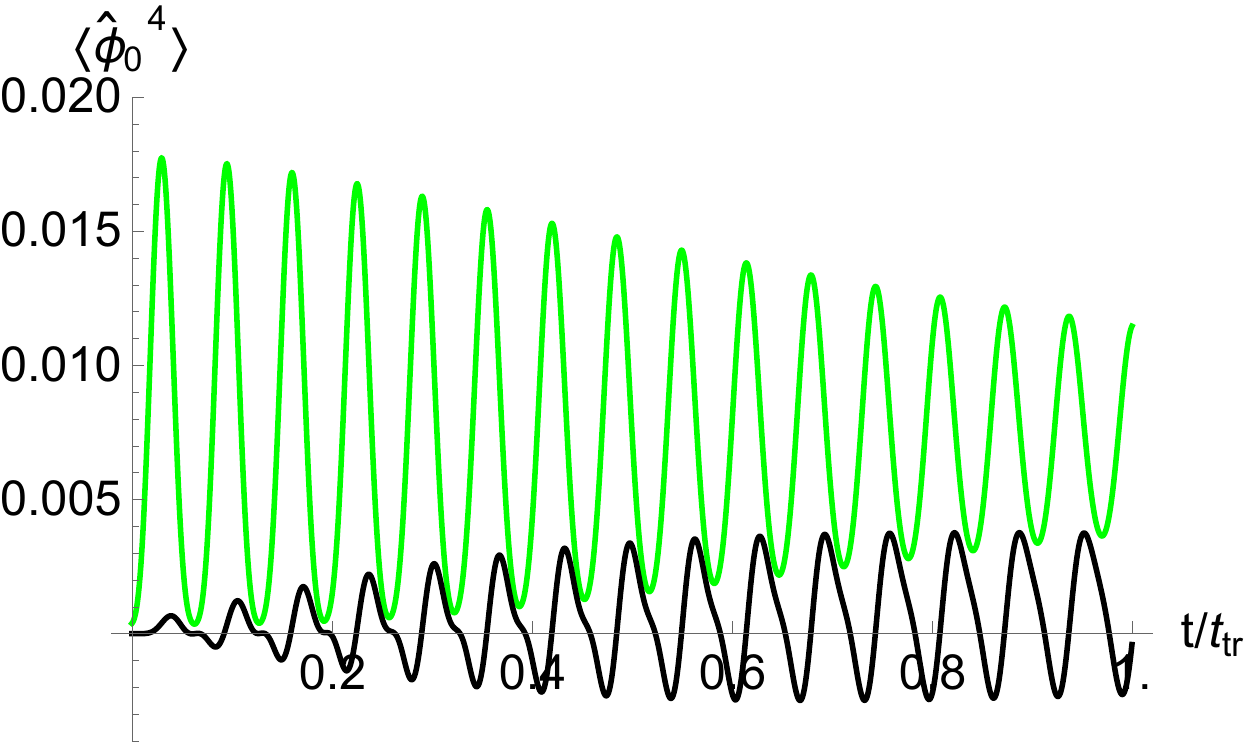}
\caption{Time-evolution of higher moments under the full cosine
potential (green lines) compared to the contributions of the higher
cumulants $\left< \hat{\phi}_{0}^{3} \right>_{c}$ and $\left<
\hat{\phi}_{0}^{4} \right>_{c}$ (black lines). The parameters are
as in Fig.~\ref{fig:QM_nu}(a).}
\label{fig:QM_nu_4}
\end{figure}

\begin{figure}[ht]
\centering
(a)\ \includegraphics[width=0.4\textwidth]{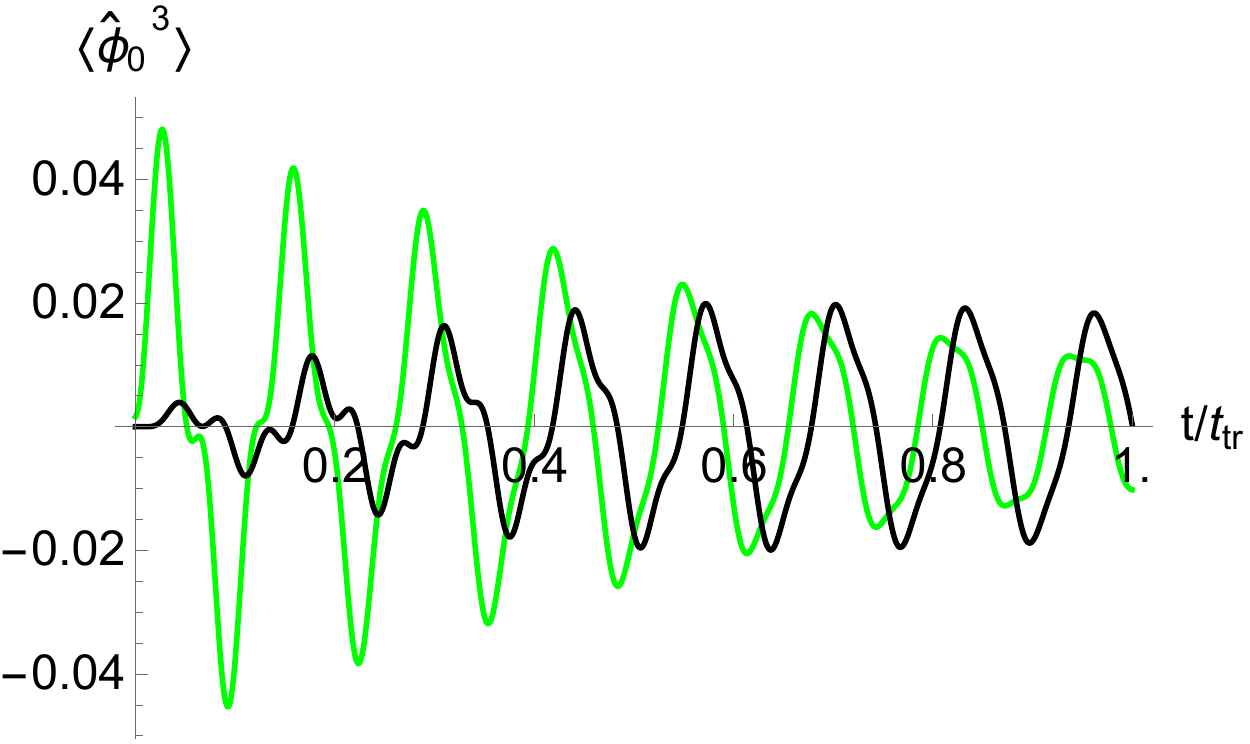}\quad
(b)\ \includegraphics[width=0.4\textwidth]{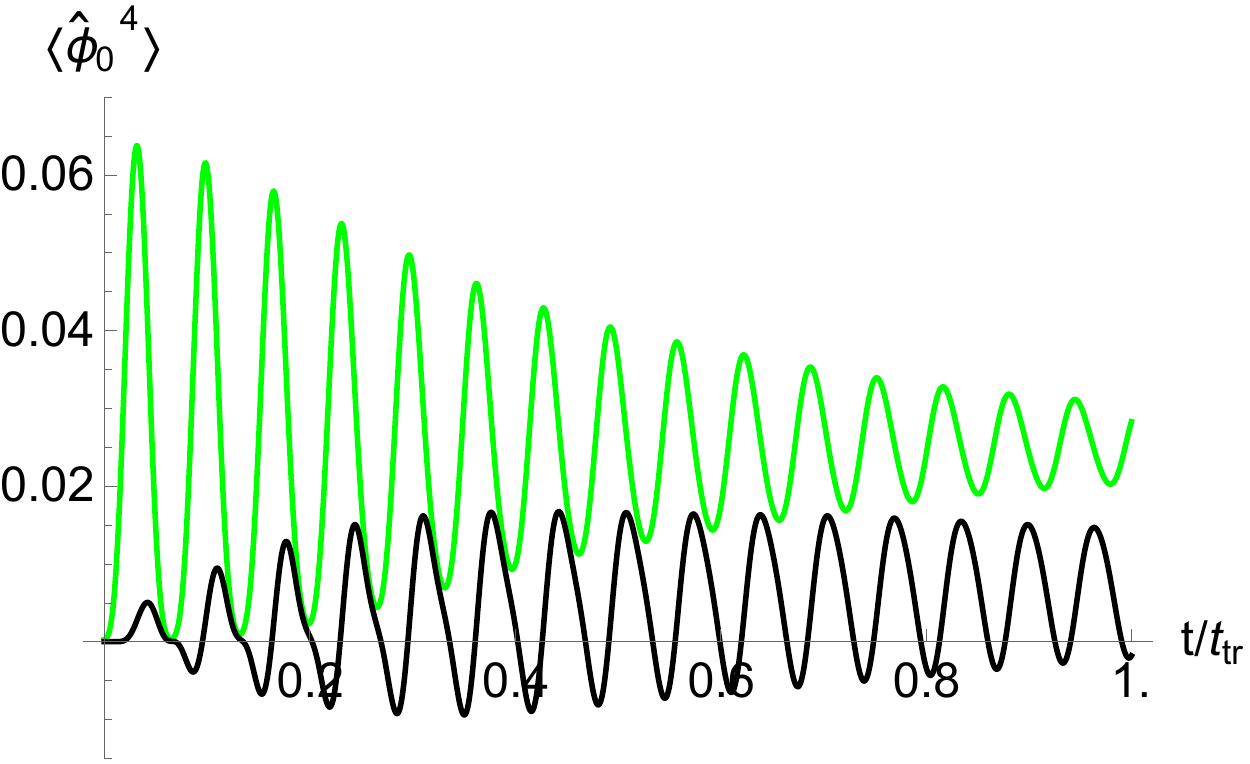}
\caption{The same as Fig.~\ref{fig:QM_nu_4}, but with squeezing
  parameter $\eta=8 \eta_{0}$.} 
\label{fig:QM_nu_8}
\end{figure}

In Fig.~\ref{fig:QM_nu} and Appendix \ref{sec:further_results_for_the_zero_mode} we compare time evolution under the
Hamiltonians $H_{\mathrm{J}}$ (green line), $H_{\mathrm{HO}}$ (red,
dotted line), and $H_{\mathrm{J}}^{\prime}$ (blue line), with
$\Phi_{0} = 0.1$ and two choices of initial state
$|\chi(0)\rangle$. We observe fast oscillations of
$\langle\hat{\phi}_0\rangle\equiv\langle\chi(t)|\hat{\phi}_0|\chi(t)\rangle$
in time with a slowly varying envelope. This envelope shrinks (Fig.~\ref{fig:QM_nu}) or expands
(Appendix \ref{sec:further_results_for_the_zero_mode}),   
depending on the initial values $\Phi_{0}$ and $\delta N_{0}$.
We observe that the amplitude modulation is more
pronounced when $\eta/\eta_{0}$ is either large or small, which
corresponds to initial states with either large phase or number
fluctuations. Such states are sensitive to the anharmonicity of the
cosine potential and their time evolution will exhibit larger
deviations from that of a simple harmonic oscillator.
We see that the SCTDHA is significantly better than the simple
harmonic approximation $H_{\rm HO}$. The SCTDHA neglects higher
connected correlations and is accurate as long as the latter are
small. The contribution of the connected correlation
functions to the expectation values of $\hat{\phi}_0^3$ and
$\hat{\phi}_0^4$ are respectively 
\begin{align}
\left< \hat{\phi}_{0}^{3} \right> &= \left< \hat{\phi}_{0}^{3} \right>_{c} + 3\left< \hat{\phi}_{0}^{2} \right>_{c} \left< \hat{\phi}_{0} \right> + \left< \hat{\phi}_{0} \right>^{3}, \\
\left< \hat{\phi}_{0}^{4} \right> &= \left< \hat{\phi}_{0}^{4} \right>_{c} + 4\left< \hat{\phi}_{0}^{3} \right>_{c} \left< \hat{\phi}_{0} \right> + 3\left< \hat{\phi}_{0}^{2} \right>_{c}^{2} + 6\left< \hat{\phi}_{0}^{2} \right>_{c} \left< \hat{\phi}_{0} \right>^{2} + \left< \hat{\phi}_{0} \right>^{4}.
\end{align}
Figs~\ref{fig:QM_nu_4} and \ref{fig:QM_nu_8} show the time
evolution of the neglected connected contributions and compare them to
the full expectation value.
By our choice of initial state the cumulants are initially zero and
then grow in time. The growth of even cumulants is inhibited by choosing
the squeezing parameter $\eta$ close to $\eta_{0}$, whereas odd
cumulants are inhibited by choosing $\Phi_{0}$ and $\delta N_{0}$
close to $0$. In our examples the cumulants remain small and
concomitantly the SCTDHA is a good approximation.

\subsection{Time-evolution in the SCTDHA for the sine-Gordon model} 
\label{sub:time_evolution_with_the_full_self_consistent_harmonic_Hamiltonian}
\begin{figure}[htbp]
\centering
(a)\ \includegraphics[width=0.4\textwidth]{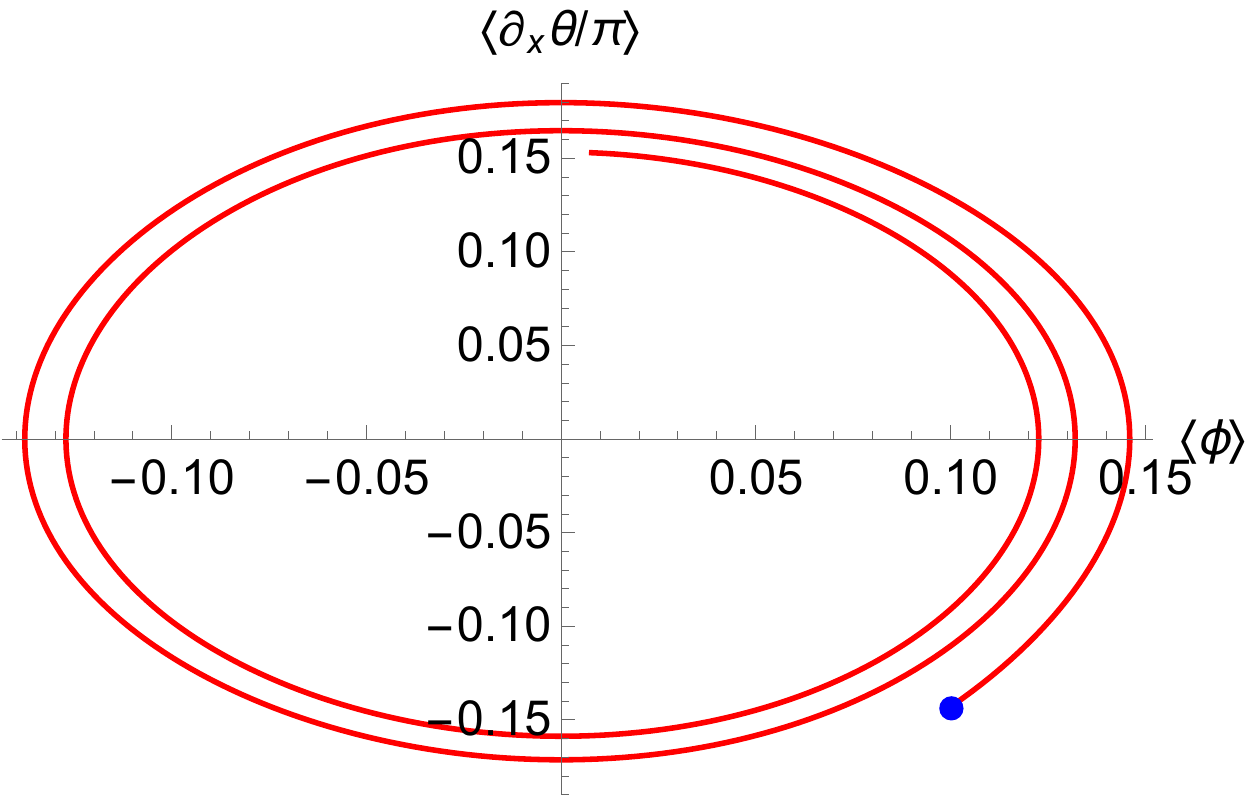}\qquad
(b)\ \includegraphics[width=0.4\textwidth]{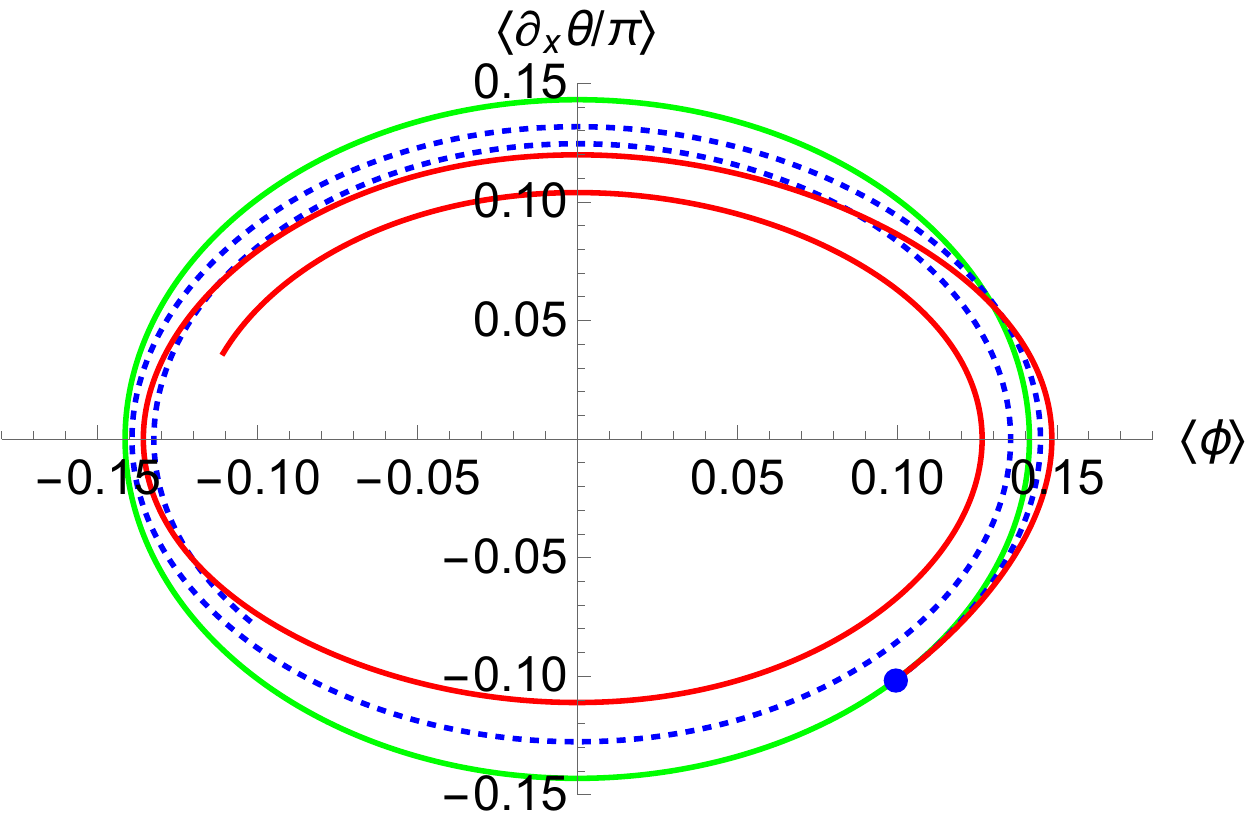}
\caption{
(a) Density-phase oscillations in the SCTDHA. The parameters are as
described in \ref{sec:note_on_the_choice_of_parameters}
and $K=28, \lambda = 0.4$, $\eta=1$. A modulation of the amplitude can
be observed, which is not present in a simple quadratic approximation.
(b) Same as (a) but with $\lambda=0.2$, $\eta=0.5$ (blue) and $\eta=1$
(red). For comparison we also show the result of a simple harmonic
approximation (green line). The modulation is seen to
increase with $\eta$. In both panels, time runs until the traversal
time $t_{\mathrm{tr}} = L/(2v)$. 
}
\label{fig:densityphase}
\end{figure}

Having tested the self-consistent harmonic approximation in the
controlled setting of single-body quantum mechanics, we now apply it
to the sine-Gordon field theory, using the formalism developed in
section \ref{sec:sine_gordon_in_self_consistent_approximation}.
Motivated by experiment we focus on the following observables:
\begin{itemize}
\item{} The one-point functions of density $\langle
\partial_x\theta(x,t)\rangle /\pi$ and phase
$\langle\phi(x,t)\rangle$. As we are restricting ourselves to
translationally invariant situations these are $x$-independent.
\item{} The full quantum mechanical probability distribution of
$\int_{-\ell/2}^{\ell/2} dx\ \sin\big(\phi(x)\big)$ 
\be
P_\ell(t,\mu)=\langle\psi_{\rm SCH}(t)|
\delta\Big(\mu-\int_{-\ell/2}^{\ell/2}dx\ \sin\big(\phi(x)\big)\ \Big)|\psi_{\rm SCH}(t)\rangle\ .
\ee
\end{itemize}
In Fig.~\ref{fig:densityphase} we show parametric plots for the time
dependence of the average density and phase in the SCTDHA for two
different choices of parameters. In a purely harmonic theory the
resulting trajectory would be closed, \emph{cf.} the green line in
Fig.~\ref{fig:densityphase}(b). In contrast the amplitude of these oscillations gets modulated in time
in the SCTDHA. We observe that these modulations become more
pronounced as the squeezing parameter $\eta$ is increased from its
ground state value.

\begin{figure}[ht]
\centering
(a)\ \includegraphics[width=0.35\textwidth]{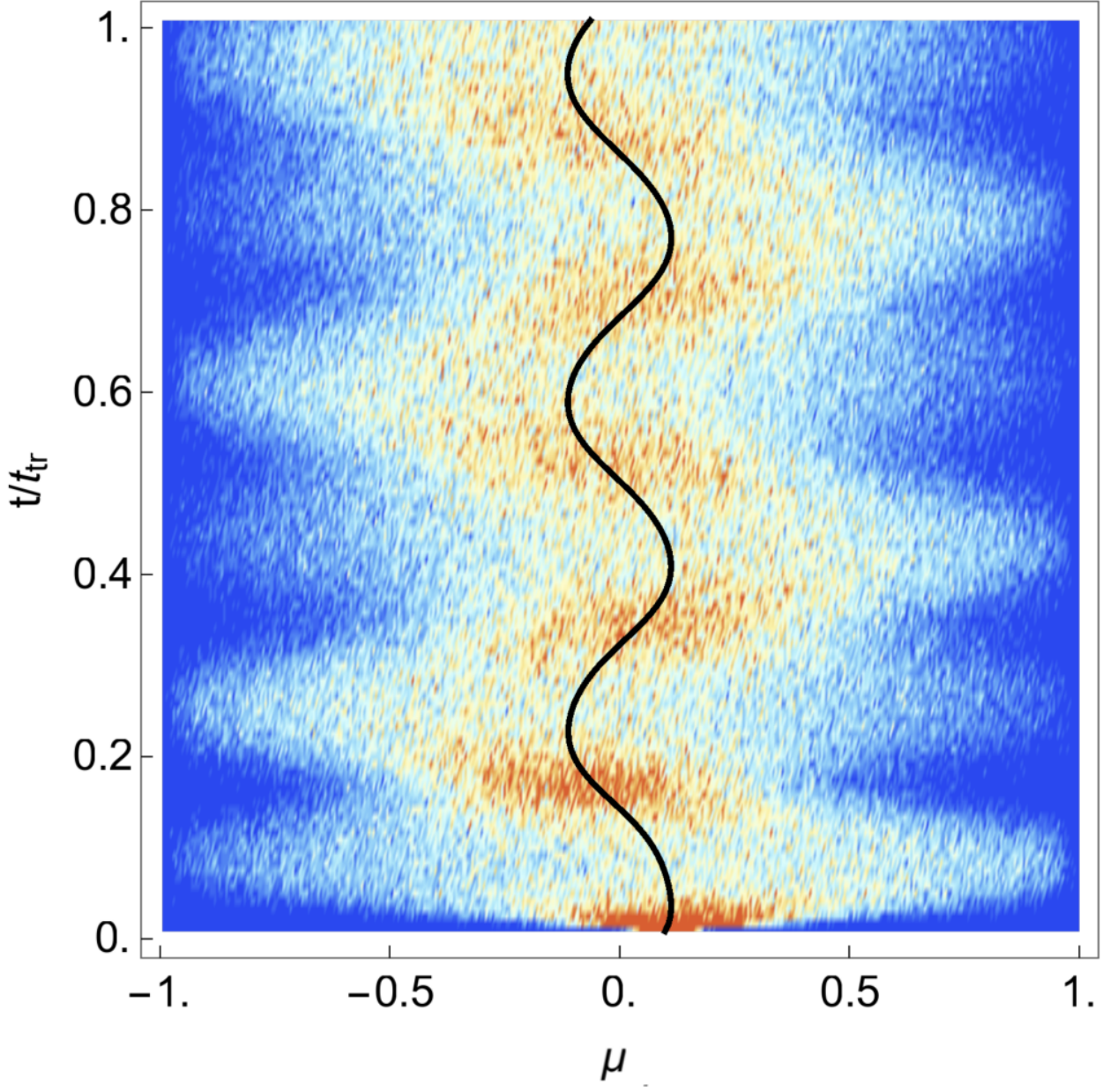}\qquad
\includegraphics[height=0.3\textwidth]{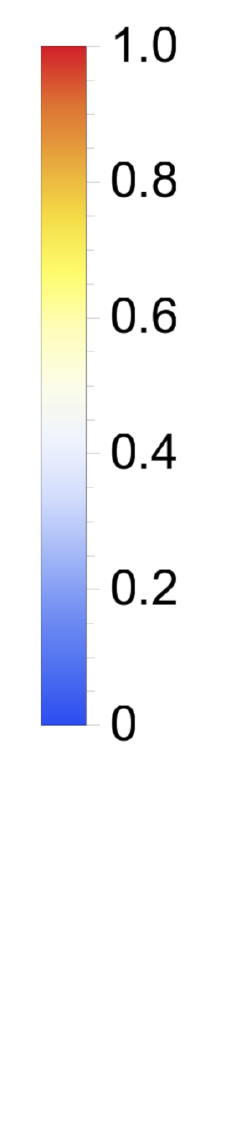}
(b)\ \includegraphics[width=0.35\textwidth]{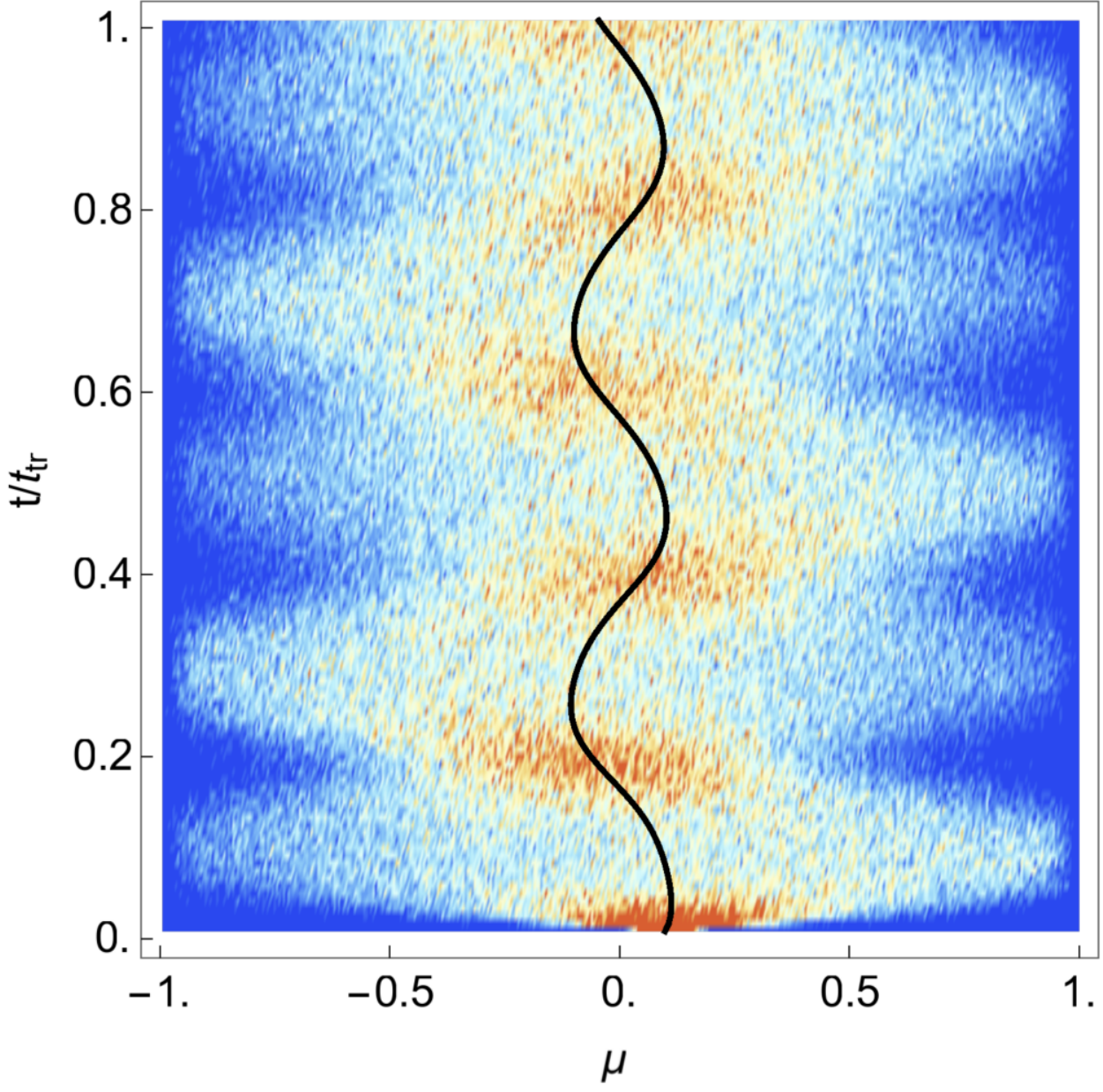}
\caption{(a) Probability distribution function $P_{\ell}(t,\mu)$ for
a very short integration length $\ell=\xi$ in a simple harmonic
approximation to the sine-Gordon model corresponding to $g=0$ and
$h=-1/2$ in (\ref{eq:fg_def}). Parameters are as described in 
\ref{sec:note_on_the_choice_of_parameters} with $K=28$ and $\lambda = 0.2$.
The black line shows the average of the
PDF. (b) Same as (a) but computed in the SCTDHA.} 
\label{fig:full_SCHA_phase_FDF}
\end{figure}

\begin{figure}[ht]
\centering
(a)\ \includegraphics[width=0.35\textwidth]{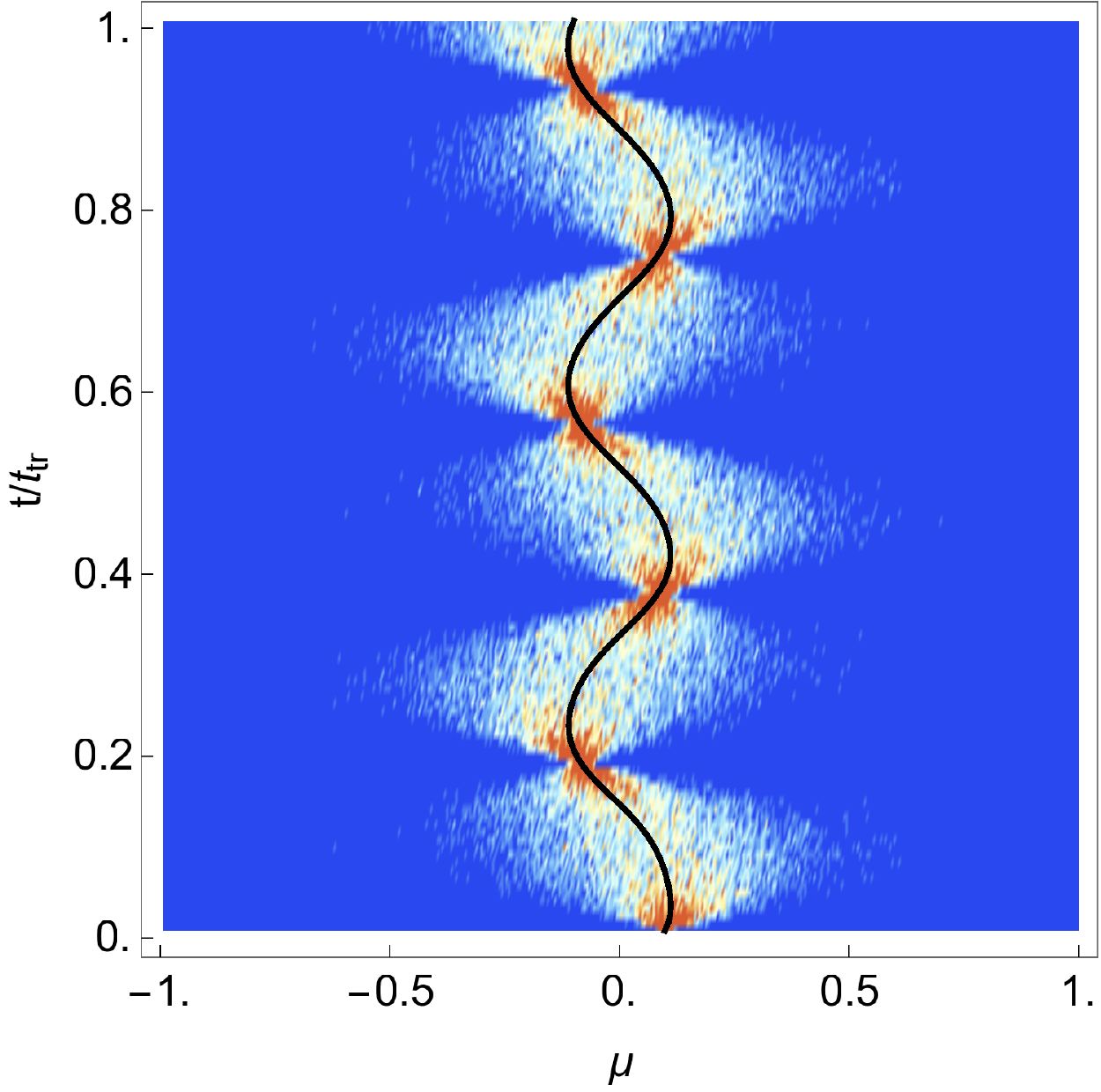}\qquad
\includegraphics[height=0.3\textwidth]{Colorbar.pdf}
(b)\ \includegraphics[width=0.35\textwidth]{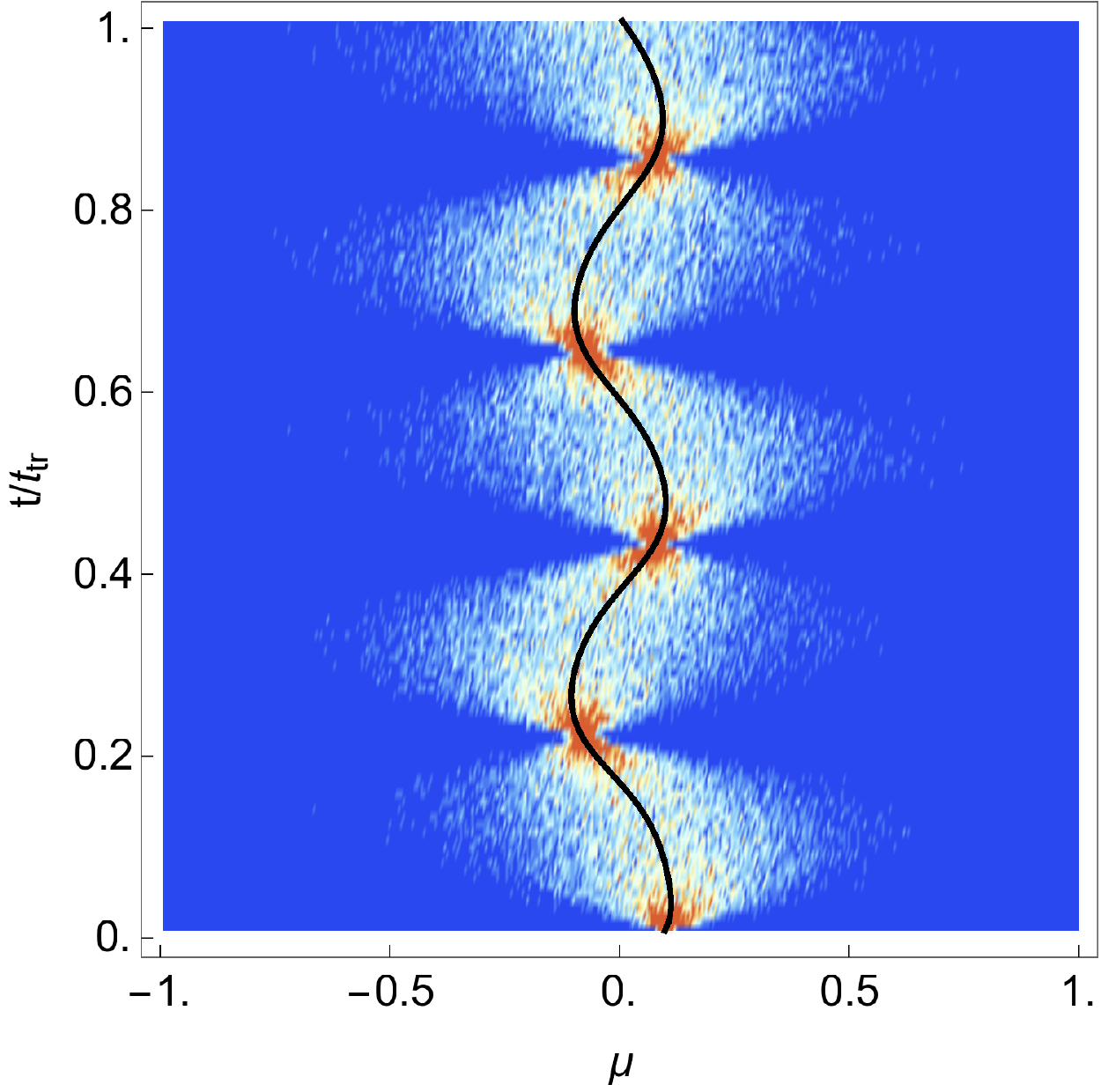}
\caption{Same as Fig.~\ref{fig:full_SCHA_phase_FDF} but with a long integration
length $\ell=L$. }
\label{fig:full_SCHA_phase_FDF_long}
\end{figure}

We now turn to the probability distribution function $P_\ell(t,\mu)$. In
recent experiments \cite{Pigneur2017} it was observed that the
variance of the probability distribution of the phase exhibits a rapid
narrowing. A detailed explanation why these experiments have access to
the probability distribution of the phase itself is given in
Ref.~\onlinecite{Nieuwkerk2018}. An important question is whether such
behaviour arises in the framework of the sine-Gordon model.
In Figs~\ref{fig:full_SCHA_phase_FDF} and
\ref{fig:full_SCHA_phase_FDF_long} we show results for $P_\ell(t,\mu)$
for two integration lengths $\ell$ obtained in the SCTDHA and a simple
harmonic approximation. Both display oscillatory behaviour in time and
no narrowing of the variance is observed. In fact, the variance in the
SCTDHA is slightly larger than the simple harmonic result. Comparing
Fig.~\ref{fig:full_SCHA_phase_FDF} to \ref{fig:full_SCHA_phase_FDF_long}
we observe that increasing the integration length leads to a narrowing of
$P_\ell(t,\mu)$. 
\section{Dynamics at the Luther-Emery (LE) point} 
\label{sec:dynamics_at_the_luther_emery_point}
The SCTDHA is expected to work best at large values of the Luttinger
parameter $K$. It is instructive to complement the large-$K$ results
presented above by exact results at the free fermion point of the
sine-Gordon model. In our conventions the LE point occurs at
$K=1/4$. Quench dynamics at the LE point has been previously
considered in Ref. \onlinecite{Iucci2009,Iucci2010} but that analysis did not
cover the class of initial states of interest to us here.
Two remarks are in order before we proceed:
\begin{itemize}
\item{} The LE point occurs at an unphysical value of $K$ as far as the
realization of the sine-Gordon model in the context of tunnel-coupled
Bose gases is concerned. In that context the Luttinger parameter
runs from $K=1/2$ (hard-core repulsion) to $K \rightarrow \infty$
(non-interacting bosons).
\item{} The SCTDHA is not expected to be a good approximation at the
LE point. We have already seen an example of this in section
\ref{sub:linear_response_above_the_ground_state_}. The fundamental
problem is that the relevant degrees of freedom at the LE point are
solitons and antisolitons and these are not captured by a harmonic
approximation. In light of this we will refrain from attempting to
apply the SCTDHA to the sine-Gordon model at $K=1/4$.
\end{itemize}  

\subsection{Mapping to free fermions} 
\label{sub:mapping_to_free_fermions}
The sine-Gordon model can be fermionized using the bosonization identities
\begin{align}
R(x) = \frac{F}{\sqrt{2 \pi \xi}}\, e^{-i \sqrt{4 \pi}
  \varphi_{R}(x)}, \;\;\;\;L(x) = \frac{\bar{F}}{\sqrt{2 \pi \xi}}\,
e^{i \sqrt{4 \pi}
  \varphi_{L}(x)}, \label{eq:bosonization_id_fermionize} 
\end{align}
where $F = \sigma_{x}, \bar{F} = \sigma_{y}$ are Klein factors and
$\varphi_{R/L}(x)$ are chiral Bose fields defined as
\be
\varphi_{R/L} = \sqrt{\frac{K}{4\pi}} \phi \pm \frac{1}{\sqrt{4\pi
    K}}\ \theta \ .
\label{eq:def_phi_left_right}
\ee
The fields defined in \fr{eq:bosonization_id_fermionize} fulfil
anticommutation relations
$\{R^{\dagger}(x),R(y)\} = \{L^{\dagger}(x),L(y)\} =
\delta(x-y)$. Expectation values are always taken with respect to the
vector $(1, 0)$ in Klein space. At the LE point the sine-Gordon Hamiltonian
(\ref{eq:Sine-Gordon_ham_full}) is equivalent to
\begin{align}
H_{\mathrm{F}} = \int_{L} dx \, \left[ i v \left( L^{\dagger}(x)
\partial_{x} L(x) - R^{\dagger}(x) \partial_{x} R(x) \right) + i \mu
\left( R^{\dagger}(x)L(x) - L^{\dagger}(x)R(x) \right)
\right],
\label{eq:ferm_ham} 
\end{align}
where $\mu  = \pi \xi J = \pi v \lambda/\xi$ and $v$, $\xi$ and $\lambda$ respectively are the physical sound velocity, coherence length and dimensionless coupling defined in section \ref{sec:Realization_by_tunnel-coupled_bose_gases}.

\subsection{Time-evolution of density and phase} 
\label{sub:time_evolution_of_density_and_phase}

\begin{figure}[htbp]
\centering
(a)
\includegraphics[width=0.4\textwidth]{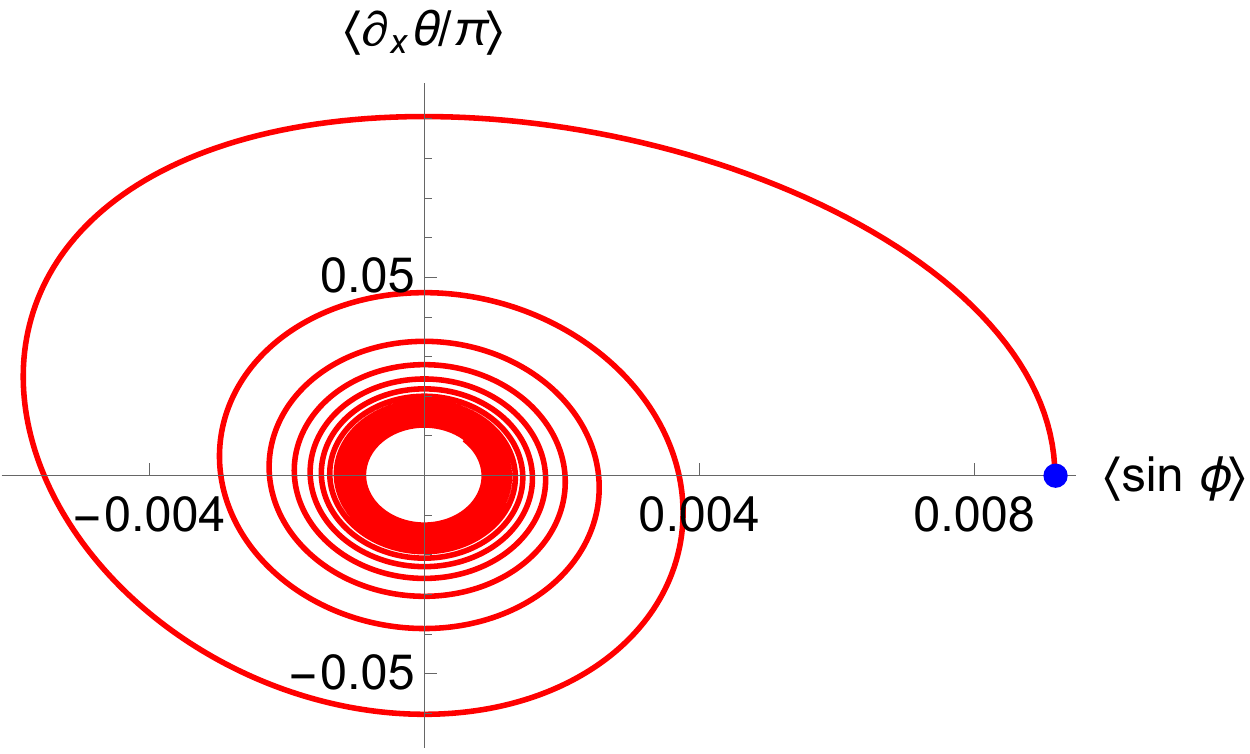}
(b)
\includegraphics[width=0.4\textwidth]{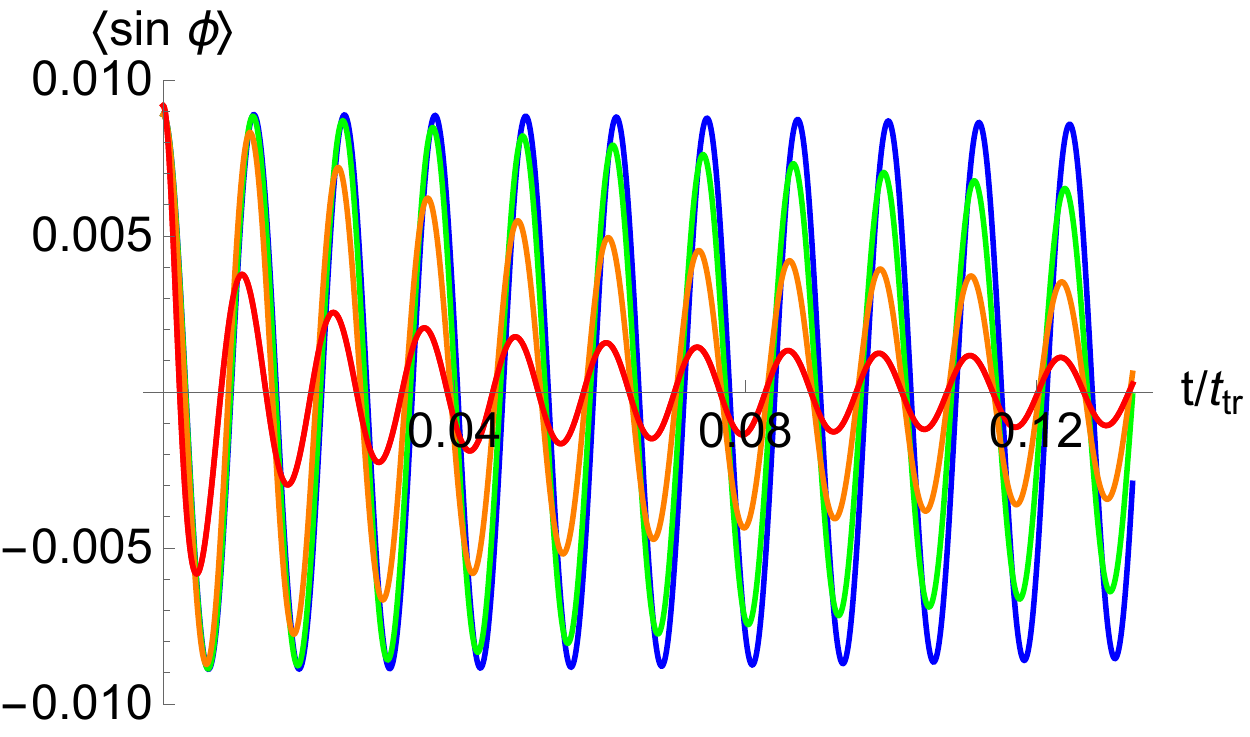}
\caption{(a) Strongly damped density-phase oscillation at the Luther-Emery
  point. Apart from taking $K=1/4$, we have used the parameters as
  reported in (\ref{sec:note_on_the_choice_of_parameters}),
  with the dimensionless coupling constant
  (\ref{eq:dimensionless_coupling}) set to $\lambda = 4$. The initial
  conditions are obtained from Eqs. (\ref{eq:init_state}) and
  (\ref{eq:alpha_def}) using the bosonization identity
  (\ref{eq:bosonization_id_fermionize}), with $\left< \delta N
  \right>  = 0$ and $\left< \phi_{0} \right> = 0.1$ at $t=0$. Due to the
  enhanced phase fluctuations at the Luther-Emery point for the state
  under consideration, the expectation value of the sine is reduced to
  $\left< \sin \phi \right> \approx 0.009$, at $t=0$. Time runs until
  the traversal time $t_{\mathrm{tr}} = L/(2v)$. (b) Oscillations of $\left< \sin \phi(0,t) \right> $ for a range of initial states. Along with the initial conditions from (a) shown in red, we plot results where $D_{\phi}(x,0)$ is a Gaussian with standard deviation $\ell = \nu \xi$, for $\nu=1$ (orange), $\nu=2$ (green) and $\nu=4$ (blue). For comparison, the initial conditions for $D_{\phi}(x,0)$ pertaining to the red line are sharply peaked around $x=0$ with standard deviation $\ell \approx 0.34 \xi$. }  
  \label{fig:Dens_Phase_Luther_Enery}
\end{figure}

Our aim is to determine the expectation values of
\bea
\sin\big(\phi(x,t)\big) &=& - \pi \xi \left[R^{\dagger}(x,t)L(x,t)+{\rm
    h.c.}\right], \nn
\frac{\partial_{x}\theta(x,t)}{\pi} - \frac{\left< \partial_{x}\theta(x,0)\right> }{\pi}  &=& \frac{1}{2} \left[
  :L^{\dagger}(x,t)L(x,t): - :R^{\dagger}(x,t)R(x,t):
  \right].
 \label{eq:bosonization_dens} 
\eea
Here products of operators at the same point are defined by means of a
point-splitting prescription  
\begin{align}
:L^{\dagger}(x)L(x): \equiv \lim_{\epsilon \rightarrow 0} \left[ L^{\dagger}(x-\epsilon)L(x+\epsilon) - \left< L^{\dagger}(x-\epsilon)L(x+\epsilon) \right>_{0} \right],
\end{align}
where $\left< \ldots \right>_{0}$ denotes the expectation value with
respect to the initial state under consideration. In order to make
some contact with our previous discussion we choose the initial state
to be $\ket{V,\varphi,\vartheta}$ in (\ref{eq:init_state}) and (\ref{eq:alpha_def}), i.e.
\be
\langle {\cal O}\rangle_0\equiv\bra{V,\varphi,\vartheta}{\cal O}\ket{V,\varphi,\vartheta}. \label{eq:init_cond_LE}
\ee
This state is translationally invariant, as is the Hamiltonian (\ref{eq:ferm_ham}), so that the expectation values of the fields (\ref{eq:bosonization_dens}) do not depend on $x$ for any $t$. To determine the time evolution of these expectation values, it is useful to define the following linear combinations of two-point functions,
\begin{align}
D_{\phi}(x,t) &\equiv \left< R^{\dagger}(x)L(0)\right>_{t} + \left< L^{\dagger}(x)R(0)\right>_{t} + \left< R^{\dagger}(0)L(x)\right>_{t} + \left< L^{\dagger}(0)R(x)\right>_{t} ,\nn
D_{\theta}(x,t) &\equiv \left< L^{\dagger}(x)L(0)\right>_{t} + \left< L^{\dagger}(0)L(x)\right>_{t} - \left< R^{\dagger}(x)R(0)\right>_{t} - \left< R^{\dagger}(x)R(0)\right>_{t},
\label{Ds}
\end{align}
which only depend on $t$ and the coordinate difference $x$ due to translational invariance. The time evolution of these functions is governed by the pair of PDE's
\begin{align}
\left( \partial_{t}^{2} - 4 v^{2} \partial_{x}^{2} + 4 \mu^{2}\right) D_{\phi}(x,t) &= 0 \ , \label{eq:LE_PDE1}\\
\partial_{t} D_{\theta}(x,t) + 2 \mu D_{\phi}(x,t)&= 0 \ , \label{eq:LE_PDE2}
\end{align}
with initial conditions (\ref{eq:init_cond_LE}) and
\begin{align}
\partial_{t} D_{\phi}(x,0) = 2 \mu D_{\theta}(x,0) - 2 v \partial_{x} \left[ \left< R^{\dagger}(x)L(0)\right>_{0} + \left< R^{\dagger}(0)L(x)\right>_{0} - \left< L^{\dagger}(x)R(0)\right>_{0} - \left< L^{\dagger}(0)R(x)\right>_{0} \right]. \label{eq:LE_PDE_IC}
\end{align}
These follow directly from the Heisenberg equation of motion (\ref{eq:Heisenberg_EOM}) with Hamiltonian (\ref{eq:ferm_ham}), combined with the translational invariance of the initial state $\ket{V,\varphi,\vartheta}$. The resulting solutions give access to the expectation values of the fields (\ref{eq:bosonization_dens}), via
\begin{eqnarray}
\left< \sin\big(\phi(0,t)\big) \right>  &=& - \frac{\pi \xi}{2} D_{\phi}(0,t), \notag \\
\frac{\left< \partial_{x}\theta(0,t) \right> }{\pi} - \frac{\left< \partial_{x}\theta(0,0)\right> }{\pi}  &=& \frac{1}{4} \left[ D_{\theta}(0,t)-D_{\theta}(0,0) \right]. \label{eq:subtr_dens}
\end{eqnarray}
The rationale for considering the particular linear combinations of
two-point functions \fr{Ds} is to ensure cut-off independence:
the initial two-point functions $\left< L^{\dagger}(x)L(0)\right>_{0}$
and $\left< R^{\dagger}(x)R(0)\right>_{0}$ diverge as $x \rightarrow
0$ in a way that depends on the UV cutoff. This cutoff-dependence enters
the equations of motion of two-point functions via the short-distance
behaviour of $D_{\theta}(x,0)$ in the initial condition
(\ref{eq:LE_PDE_IC}). To eliminate this dependence we restrict
ourselves to initial states $\ket{V,\varphi,\vartheta}$ for which
$\left< \delta N \right>_{0} = 0$. For such states we have
\begin{align}
\left< L^{\dagger}(0)L(x)\right>_{0} = - \left< L^{\dagger}(x)L(0)\right>_{0}, \;\;\;\;\; \left< R^{\dagger}(0)R(x)\right>_{0} = - \left< R^{\dagger}(x)R(0)\right>_{0},
\end{align}
which implies that $D_{\theta}(x,0)=0$ and renders the initial
condition (\ref{eq:LE_PDE_IC}), and hence $D_\phi(x,t)$ and $D_\theta(x,t)$,
cutoff-independent. In Fig.~\ref{fig:Dens_Phase_Luther_Enery}(a) we
present results obtained by numerically integrating the system of PDEs
for parameters as in \ref{sec:note_on_the_choice_of_parameters} with
$K=1/4$. In contrast to the modest amplitude modulations encountered
for larger $K$ in Section
\ref{sub:time_evolution_with_the_full_self_consistent_harmonic_Hamiltonian}
a strong damping of the density-phase oscillation is observed. The
origin of the damping is simple dephasing.

To shed some more light on the time-dependence of the observed
dephasing behaviour we have considered other initial states. In
Fig.~\ref{fig:Dens_Phase_Luther_Enery}(b) we compare the time
evolution of $\langle\sin\phi(0,t)\rangle$ shown in 
Fig.~\ref{fig:Dens_Phase_Luther_Enery}(a) to that corresponding to
initial states characterized by initial conditions
\be
D_{\phi}(x,0)=- \frac{2}{\pi \xi} \braket{\sin \phi(0,0)} e^{-x^2/2(\nu\xi)^2}\ ,\quad \nu=1,2,4. \label{eq:initial_Dphi}
\ee
As the length scale $\ell=\nu\xi$ set by $D_{\phi}(x,0)$ is increased,
the dephasing is seen to disappear. This can be understood by noting
that (\ref{eq:LE_PDE1}) is simply a Klein-Gordon equation with
dispersion relation $\omega_{k} = 2v \sqrt{k^{2} + (\pi\lambda/\xi)^{2} }$.
A wave packet $D_{\phi}(x,0)$ of initial width $\ell$ that is
initially localized around the origin will disperse. The quantity of
interest, $\langle\sin\phi(0,t)\rangle$, corresponds to the magnitude of
$D_{\phi}(0,t)$, i.e. the part of the wave packet that remains at the
origin. If the initial width of $D_{\phi}(x,0)$ is much smaller than
the inverse gap, $\ell \ll \xi/(\pi \lambda)$, the initial time
evolution will be dominated by the large-$k$ Fourier modes where the
dispersion is approximately linear. This causes the wave packet to
essentially separate into parts that propagate ballistically with
velocities $\pm 2v$. This leaves only a small weight near the origin
and leads to a rapid decrease of $\left< \sin \phi(0,t) \right>$.
In contrast, the short-time evolution of wave packets with widths that
far exceed the inverse gap $\ell \gg \xi/(\pi \lambda)$ is dominated by
Fourier modes at small $k$, where the group velocity $\frac{\partial
  \omega}{\partial k} \ll v$ becomes very small. This results in a
very slow evolution so that the weight at $x=0$ is not substantially
reduced for long times. The behaviour shown in
Fig.~\ref{fig:Dens_Phase_Luther_Enery}(b) is in complete agreement
with these expectations.

The above observations are quantified by going over to momentum space
\begin{align}
R(x) = \frac{1}{\sqrt{L}} \sum_{k} e^{i k x}\ a_k\ ,\quad
L(x) = \frac{1}{\sqrt{L}} \sum_{k} e^{i k x}\ b_k\ .
\label{eq:mode_expansions_fermions}
\end{align}
The Hamiltonian is expressed in terms of the modes as
\begin{align}
H = \sum_{k} \left( v k \left[ a_{k}^{\dagger} a^{\phdag}_{k} - b_{k}^{\dagger} b^{\phdag}_{k} \right] + i \mu \left[ a_{k}^{\dagger} b^{\phdag}_{k} - b_{k}^{\dagger} a^{\phdag}_{k} \right] \right).
\end{align}
The solution to the equations of motion is
\begin{align}
\begin{pmatrix}
  a_{k}(t) \\
  b_{k}(t)
\end{pmatrix} = \begin{pmatrix}
  \cos(\omega_{k} t) - i \sin(\omega_{k} t) \cos(2 \gamma_{k}) & \sin(\omega_{k} t) \sin(2\gamma_{k})\\
  -\sin(\omega_{k} t) \sin(2\gamma_{k}) & \cos(\omega_{k} t) + i \sin(\omega_{k} t) \cos(2 \gamma_{k})
\end{pmatrix} \begin{pmatrix}
  a_{k} \\
  b_{k}
\end{pmatrix},
\end{align}
where
\be
\sin(2 \gamma_{k}) = \frac{\mu}{\omega_{k}}\ ,\quad
\cos(2 \gamma_{k}) = \frac{v k}{\omega_{k}}\ ,\quad
\omega_{k} = \mathrm{sgn}(k) \sqrt{(v k)^{2} + \mu^{2}}.
\ee
Two-point functions of Fermi fields can be straightforwardly
calculated. Using the bosonization identities
(\ref{eq:def_phi_left_right}), (\ref{eq:bosonization_id_fermionize})
they can then be related to expectation values of fields in the sine-Gordon model.
Specializing to translationally invariant initial states with initial
condition $\left< \partial_{x}\theta(x)\right>_{0} = 0$ we find
\bea
\frac{1}{\pi}\left< \partial_{x}\theta(x,t)\right>_{0}  &=&
-\frac{1}{L} \sum_{k} \frac{\mu }{\omega_{k}}\sin(2\omega_{k}t)\
\mathrm{Re}\braket{a_{k}^{\dagger} b_{k}^{\phdag}},\nn
\left< \sin \phi(x,t) \right>_{0} &=& -
\frac{1}{L}\sum_{k}2\pi\xi \cos(2\omega_{k} t)\ \mathrm{Re}
 \braket{a_{k}^{\dagger} b_{k}^{\phdag}}. \label{eq:sin_phi_sol_LE}
\eea
The form of Eq. (\ref{eq:sin_phi_sol_LE}) allows us to relate the
origin of the observed dephasing to properties of the initial state. 
If the weights $|\braket{a_{k}^{\dagger} b_{k}^{\phdag}}|$ are
concentrated in the small momentum region one can approximate
\begin{align}
\left< \sin \phi(x,t) \right>_{0} \approx -
\frac{\cos(2 \mu t)\ }{L}\sum_{k}2\pi\xi \ \mathrm{Re}
 \braket{a_{k}^{\dagger} b_{k}^{\phdag}},
\end{align}
showing undamped oscillations at frequency $2 \mu$ over a large
time-window. On the other hand, if the weights are concentrated at
large momenta strong dephasing sets in immediately.

\section{Conclusions} 
\label{sec:conclusions}
We have implemented a self-consistent time-dependent approximation for
the quantum sine-Gordon model out of equilibrium. The approximation
incorporates anharmonic effects of the cosine potential in a
time-dependent manner by reducing higher-order fluctuations of the
phase field to time-dependent mean field coefficients in the
Hamiltonian. This leads to a time-dependent non-interacting Hamiltonian that
can be analyzed by standard methods. Its simple structure allows for
the calculation of multi-point correlation functions and full
quantum mechanical probability distribution functions of some
observables out of equilibrium.

As an application, we have considered tunnel-coupled, coherently split
Bose-gases with an initial density- and phase offset. We found that
expectation values of the density and phase exhibit oscillatory
behaviour with amplitudes that are modulated in time. Such modulations
are not observed in a simple harmonic approximation and arise from the
anharmonicity of the cosine potential. These findings are of interest in relation to recent experiments by
the Vienna group \cite{Pigneur2017}, where qualitatively similar
behaviour was observed. However, the SCTDHA does not provide a
quantitative explanation of the experimental findings. Moreover, the
experiments show a rapid narrowing of the probability distribution of
the phase, in contrast to what we find in the SCTDHA. Our results are
in accord with recent numerical studies \cite{Horvath2018} and
suggest that a simple sine-Gordon model is insufficient for
describing the experiments. 

Interestingly, an exact calculation at the free fermion point of the
sine-Gordon model shows strong  damping of oscillations, rather than
the modest modulations encountered for weak interactions.
While this is not applicable to experiments on tunnel-coupled bosons
since the Luther-Emery point occurs at an unphysical value of the
Luttinger parameter, it suggests that stronger interactions in the
sine-Gordon model lead to an enhancement of the damping effects.

Our self-consistent method is very general and can in particular be
applied to inhomogeneous situations. In a forthcoming publication we use it to
analyze interactions between the symmetric and antisymmetric sectors
in tunnel-coupled Bose gases and consider situations that are not
translationally invariant\cite{NieuwkerkPrep}. The question whether
such extensions of the theory lead to a better match with experiment will 
also be addressed there.
\acknowledgments
We are grateful to J\"{o}rg Schmiedmayer and Marine Pigneur for
stimulating discussions and to the Erwin Schr\"odinger
International Institute for Mathematics and Physics for hospitality
and support during the programme on \emph{Quantum Paths}. This work
was supported by the EPSRC under grant EP/N01930X and YDvN is
supported by the Merton College Buckee Scholarship and the VSB and
Muller Foundations. 

\appendix
\section{Initial states}
\label{app:Wick}
In this Appendix we construct a class of initial states in which a
Wick's theorem holds.
Let $b_j$ be the annihilation operators in the mode expansion of the
Bose field and consider canonical transformations of the form
\be
b^{\phdag}_j=A^{\phdag}_{jk}a^{\phdag}_k+B^{\phdag}_{kj}a^\dagger_k+v_j\ ,
\ee
where $[a^{\phdag}_k,a^\dagger_k]=\delta^{\phdag}_{j,k}$ and
\be
a_j|i\rangle=0.
\ee
For the transformation to be canonical we require
\be
AB=(AB)^T\ ,\quad
AA^\dagger-(B^\dagger B)^T=\boldsymbol{1}.
\ee
By construction we have a Wick's theorem in the state $|i\rangle$ and
the relevant one and two-point functions are
\bea
\langle i|b_j|i\rangle&=&v_j\ ,\nn
\langle i|b_kb_p|i\rangle
-\langle i|b_k|i\rangle\langle i|b_p|i\rangle&=&(AB)_{kp}\ ,\nn
\langle i|b^{\phdag}_k b^\dagger_p|i\rangle-
\langle i|b^{\phdag}_k|i\rangle\langle i|b_p^\dagger|i\rangle
&=&(AA^\dagger)^{\phdag}_{kp}.
\eea

\section{Joint Distribution Functions for the phase operator} 
\label{sec:joint_distribution_functions_for_the_phase_operator}
The goal of this Appendix is to compute full distribution functions
for the real and imaginary parts of the following operator
\begin{align}
\hat{\mathcal{O}}_{\ell}= \int_{-\ell/2}^{\ell/2} dx\, e^{i
  \hat{\phi}(x,t)}, \label{eq:FCS_complex_operator_app} 
\end{align}
where the time evolution is calculated in the SCTDHA. The real and
imaginary parts of $\hat{\mathcal{O}}_{\ell}$ are Hermitian
and their respective measurement outcomes can be described by a joint
PDF $F_{\ell}(t,a,b)$, which gives the probability density of
simultaneously measuring the eigenvalue $a$ for $\mathrm{Re}
(\hat{\mathcal{O}}_{\ell})$ and the eigenvalue $b$ for $\mathrm{Im}
(\hat{\mathcal{O}}_{\ell})$ at time $t$. Once the joint PDF is known,
expectation values of analytic functions
$g(\mathrm{Re}(\hat{\mathcal{O}}_{l}) ,
\mathrm{Im}(\hat{\mathcal{O}}_{l}))$ can be computed via  
\begin{align}
\left<g\left(\mathrm{Re} (\hat{\mathcal{O}}_{\ell}), \mathrm{Im} (\hat{\mathcal{O}}_{\ell})\right) \right>_{t} = \iint d a\, d b\, F_{\ell}(t,a,b) g(a, b).
\end{align}
Expanding the approach of Ref.~\onlinecite{Kitagawa2011}, this
Appendix presents a computation of the PDF $F_{\ell}(t,a,b)$, by
determining the generic $(m,n)^{\mathrm{th}}$ moment $\left<\left(
\mathrm{Re} (\mathcal{O}_{\ell}) \right)^{m}\left( \mathrm{Im}
(\mathcal{O}_{\ell}) \right)^{n}\right>$, and comparing it to the
definition 
\begin{align}
M_{mn}(\ell,t)=\left<\left( \mathrm{Re} (\hat{\mathcal{O}}_{\ell}) \right)^{m}\left( \mathrm{Im} (\hat{\mathcal{O}}_{\ell}) \right)^{n}\right>_{t} = \iint d a\, d b\, F_{\ell}(t,a,b) a^{m} b^{n}, \label{eq:mnth_moment_general}
\end{align}
from which $F_{\ell}(t,a,b)$ is then extracted.
Expanding sines and cosines in terms of complex exponentials we have
\begin{align}
M_{mn}(\ell,t)
&= \left(\frac{1}{2}\right)^{m}
\left(\frac{1}{2i}\right)^{n} \sum_{\{s_{j}=\pm 1\}}
\left(\prod_{j=m+1}^{m+n}s_{j} \right) \left( \prod_{k=1}^{m+n}
\int_{-l/2}^{l/2} dx_{k} \right) \left< \prod_{l=1}^{m+n} e^{i s_{l}
  \phi_{a}(x_{l},t)} \right>. \label{eq:generic_moment_factorized} 
\end{align}
We recall that the mode expansion (\ref{eq:phi_time_evol}) for the
time evolved Bose field has the form
\begin{align}
\phi(x,t) &= \left< \phi(0,t) \right>  + \sum_{j} u_{j} e^{i q_{j} x} \left( Q^{\phdag}_{j}(t) a^{\phdag}_{j} - Q^{*\phdag}_{-j}(t) a^{\dagger}_{-j} \right)\ ,\label{eq:phi_time_evol_app}
\end{align}
where $a_j$ annihilate the initial state and with $\left< \phi(0,t)
\right>$ given by (\ref{eq:phi_exp_val}). To proceed we define
functions  
\begin{align}
w_{k}(\vect{x}) = \sum_{j=1}^{m+n} s_{j} u_{k}e^{i q_{k} x_{j}} . \label{eq:w_vect_app}
\end{align}
The expectation value (\ref{eq:generic_moment_factorized}) in the initial state can be
expressed in the form
\begin{align}
\left< \prod_{j=1}^{m+n} e^{i s_{j} \phi_{a}(x_{j},t) } \right> &= e^{i \sum_{j=1} s_{j} \left< \phi(0,t) \right>}\left< e^{i \sum_{j} w_{j}(\vect{x}) \left( Q^{\phdag}_{j}(t) a^{\phdag}_{j} - Q^{*\phdag}_{-j}(t) a^{\dagger}_{-j} \right)}\right> \\
&= e^{i \sum_{j=1} s_{j} \left< \phi(0,t) \right>} e^{- \frac{1}{2} \sum_{j} w^{\vphantom{*}}_{j}(\vect{x})w^{*}_{j}(\vect{x}) \left| Q_{j}(t) \right|^{2}}\ . \label{eq:vertex_exp_product}
\end{align}
The first exponent on the right-hand side of
(\ref{eq:vertex_exp_product}) contains products of expressions
involving different coordinates $x_{i}$ and $x_{j}$ with $i \neq
j$. This means that the integrals in
(\ref{eq:generic_moment_factorized}) over the coordinates $x_{j}$
cannot be separately carried out. We therefore perform a
Hubbard-Stratonovich transformation based on the identity
\begin{align}
e^{-\frac{q}{2}u^{2}} = \frac{1}{\sqrt{2 \pi q}} \int d z e^{-
  \frac{1}{2 q}z^{2}}e^{-i u z}\ . 
\end{align}
This gives
\begin{align}
e^{- \frac{1}{2} \sum_{j}
  w^{\vphantom{*}}_{j}(\vect{x})w^{*}_{j}(\vect{x}) \left| Q_{j}(t)
  \right|^{2}} = 
\int_{-\infty}^{\infty} d \alpha_{j} \int_{-\infty}^{\infty} d
\beta_{j} \frac{e^{- \frac{1}{2}\left| Q_{j}(t) \right|^{-2} \left(
  \alpha^{2}_{j} + \beta^{2}_{j} \right)  }}{2 \pi \left| Q_{j}(t) \right|^{2}}
 e^{-i \alpha_{j}
  \mathrm{Re} w_{j}(\vect{x}) - i \beta_{j} \mathrm{Im}
  w_{j}(\vect{x})}\ . \label{eq:after_HS} 
\end{align}
Substituting (\ref{eq:vertex_exp_product}), (\ref{eq:after_HS}) into
(\ref{eq:generic_moment_factorized}), we obtain
\begin{align}
\begin{split}
M_{mn}(\ell,t)= \left(\frac{1}{2}\right)^{m}
  \left(\frac{1}{2i}\right)^{n}  \sum_{\{s_{l}\}}
  \left(\prod_{l=m+1}^{m+n}s_{l}
  \right)\int_{-\infty}^\infty 
d\boldsymbol{\alpha} d\boldsymbol{\beta} & \int_{-\ell/2}^{\ell/2} d\boldsymbol{x}
\prod_j\frac{e^{- \frac{1}{2}\left| Q_{j}(t) \right|^{-2} \left(
    \alpha^{2}_{j} + \beta^{2}_{j} \right) } e^{i s_{j} \left<
    \phi(0,t) \right>}}{2 \pi \left| Q_{j}(t) \right|^{2}}\\
\times &\ \exp\bigg(-i \sum_{k} \left(  \alpha_{k} \mathrm{Re} w_{k}(\vect{x}) +
  \beta_{k} \mathrm{Im} w_{k}(\vect{x}) \right) \bigg)\ .
\label{eq:FCS_phi_operator_HS} 
\end{split}
\end{align}
Reinserting the vector $w_{k}(\vect{x})$ from Eq. (\ref{eq:w_vect_app}) and bringing the sum over signs $s_{l}$ within the product, this simplifies to
\begin{align}
\begin{split}
M_{mn}(\ell,t)&= \int_{-\infty}^\infty d\boldsymbol{\alpha}
d\boldsymbol{\beta}\prod_j\frac{e^{- \frac{1}{2}\left| Q_{j}(t) \right|^{-2} \left(
    \alpha^{2}_{j} + \beta^{2}_{j} \right) } }{2 \pi \left| Q_{j}(t) \right|^{2}}\\
&\times\quad \left( \int_{-\ell/2}^{\ell/2} dx \cos \left( \Phi(x,t,\boldsymbol{\alpha}, \boldsymbol{\beta}) \right) \right)^{m} \left( \int_{-\ell/2}^{\ell/2} dx \sin \left( \Phi(x,t,\boldsymbol{\alpha}, \boldsymbol{\beta}) \right) \right)^{n}, \label{eq:FCS_phi_operator_final}
\end{split}
\end{align}
where we have defined
\begin{align}
\Phi(x,t,\boldsymbol{\alpha}, \boldsymbol{\beta}) =  \left< \phi(0,t)
\right> - \sum_{j} |u_{j}| \Big( \alpha_{j} \cos \left( p_{j} x
\right) + \beta_{j} \sin \left( p_{j} x \right)
\Big). \label{eq:phi_fun_general_app} 
\end{align}
Comparing (\ref{eq:FCS_phi_operator_final}) to the definition of the
joint PDF in (\ref{eq:mnth_moment_general}) gives the desired
expression for the joint PDF 
\begin{align}
F_{\ell}(t,a,b) &= \int_{-\infty}^\infty d\boldsymbol{\alpha}
d\boldsymbol{\beta}\prod_j\frac{e^{- \frac{1}{2}\left| Q_{j}(t) \right|^{-2} \left(
    \alpha^{2}_{j} + \beta^{2}_{j} \right) } }{2 \pi \left| Q_{j}(t) \right|^{2}}\\
&\times\quad \delta\bigg(a -  \int_{-\ell/2}^{\ell/2} dx \cos \left(
\Phi(x,t,\boldsymbol{\alpha}, \boldsymbol{\beta}) \right) \bigg)\
\delta\bigg(b -  \int_{-\ell/2}^{\ell/2} dx \sin \left(
\Phi(x,t,\boldsymbol{\alpha}, \boldsymbol{\beta}) \right) \bigg).
\label{eq:FDF_result_app}
\end{align}
By integrating out the variables $a$ or $b$ in this final expression, PDF's of the imaginary and real parts of $\hat{\mathcal{O}}_{\ell}$ can be obtained, respectively. Furthermore, the two-point function (\ref{eq:phase_vertex_2pt}) immediately follows from Eq. (\ref{eq:vertex_exp_product}) by replacing the vector $w_{j}(\vect{x})$ in Eq. (\ref{eq:w_vect_app}) by $\tilde{w}_{j}(x) = u_{j} \left( \sigma e^{i q_{j} x} + \tau \right)$.


\section{Further plots for the zero mode} 
\label{sec:further_results_for_the_zero_mode}

\begin{figure}[htbp]
	\centering
	(a)\includegraphics[width=0.4\textwidth]{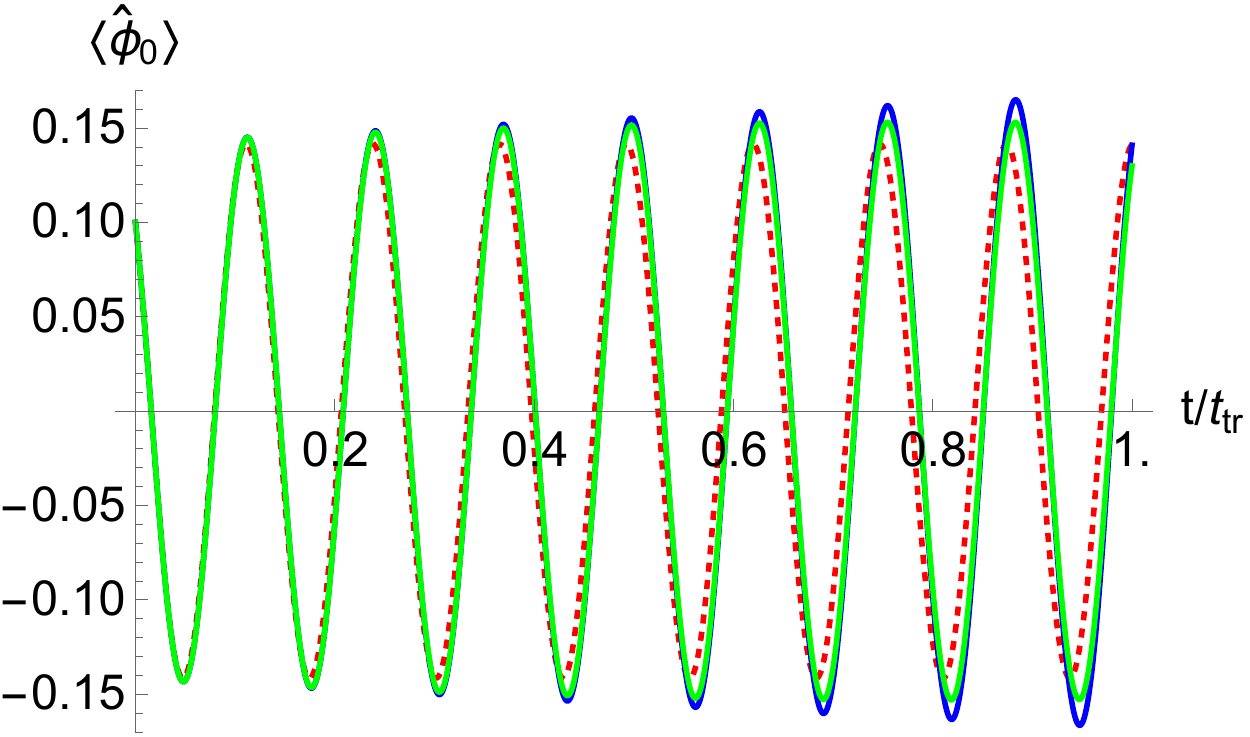}
	(b)\includegraphics[width=0.4\textwidth]{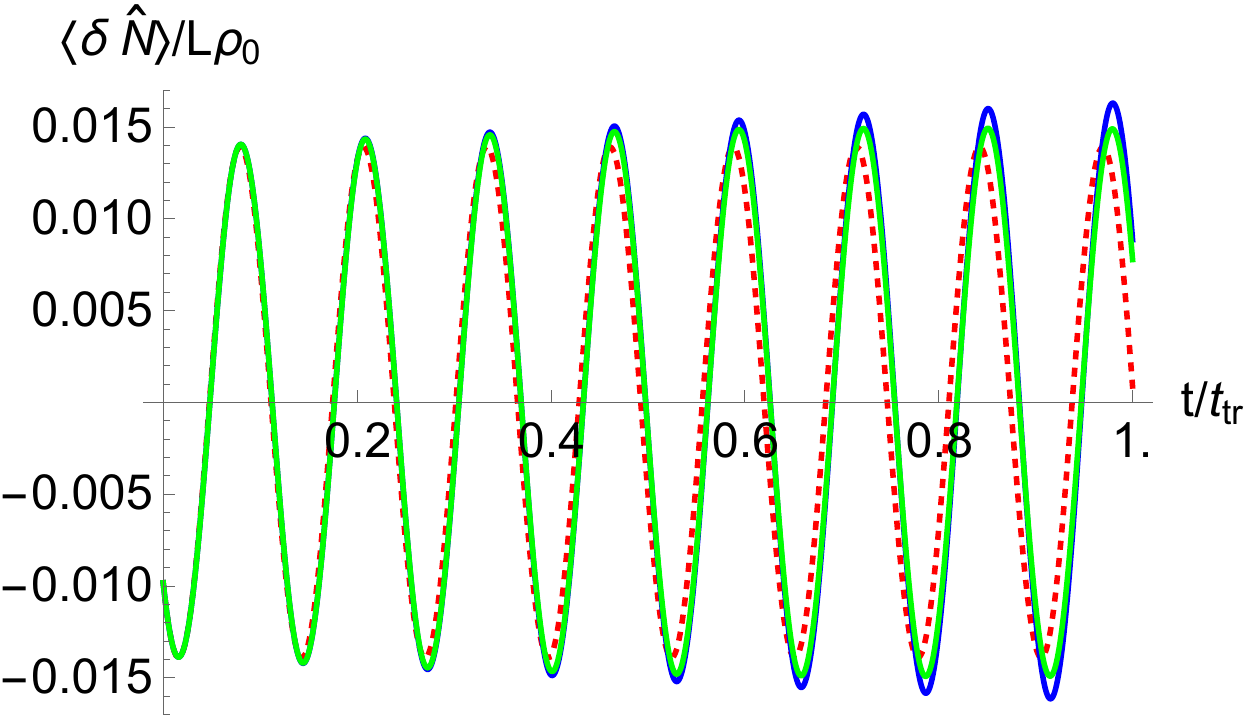}
	\caption{Comparison between time evolution with $H_{\mathrm{J}}$ (green, Eq. (\ref{eq:HJ})), the SCTDHA via $H_{\mathrm{J}}^{\prime}$ (blue, Eq. (\ref{eq:HJ_prime})) and the fully quadratic Hamiltonian $H_{\mathrm{HO}}$ (red, Eq. (\ref{eq:HHO})). Both the zero mode of the phase (a) and its conjugate variable (b) are displayed. All parameters, including the time scale, are as in Fig.~\ref{fig:QM_nu}(a), except that the sign of the initial value $\delta N_{0}$ is reversed.}
	\label{fig:envelope_increase}
\end{figure}

We here present some additional plots for the zero mode, as governed
by the quantum mechanics problem described in
Sec. \ref{sub:time_evolution_of_the_zero_mode}. In particular, we show
that the weak damping observed in Fig.~\ref{fig:QM_nu} is not the only
behavior found in the framework of the SCTDHA. A change in the initial
conditions can cause the oscillation amplitude to increase, rather
than decrease, as shown in Fig.~\ref{fig:envelope_increase}. This
widening of the envelope is particularly pronounced in the SCTDHA
result. The exact solution soon reverts to weakly damped behaviour,
though the time scale for this damping is much longer than that
observed in
Ref.~\onlinecite{Pigneur2017}. Fig.~\ref{fig:envelope_increase}
presents results for $K=1$, which corresponds to a regime in which the
quadratic approximation to $H_{\mathrm{J}}$ (Eq. (\ref{eq:HJ}))
strongly deviates from the exact solution. In contrast, there is a
reasonable correspondence between SCTDHA and exact results for the
first few periods of oscillation. For larger values of $K$, this
correspondence improves. 

\end{document}